\documentclass[12pt,preprint]{aastex}

\newcommand{\fe}{[Fe/H]}

\shorttitle{M31's Undisturbed Disk}
\shortauthors{Morrison et al.}

\begin{document}


\title{M31's Undisturbed Thin Disk of Globular Clusters}


\author{Heather L. Morrison\footnote{Cottrell Scholar of Research
Corporation and NSF CAREER fellow}}
\affil{Department of Astronomy\footnote{and Department of Physics}, 
Case Western Reserve University, Cleveland OH 44106-7215 
\\ electronic mail: heather@vegemite.astr.cwru.edu}

\author{Paul Harding}
\affil{Department of Astronomy, 
Case Western Reserve University, Cleveland OH 44106-7215 
\\electronic mail: harding@dropbear.astr.cwru.edu}

\author{Kathy Perrett}
\affil{Department of Astronomy and Astrophysics, University of
Toronto, Toronto, ON, Canada M5S 3H8
\\electronic mail: perrett@astro.utoronto.ca}

\and

\author{Denise Hurley-Keller\footnote{NSF AAP fellow}}
\affil{Department of Astronomy,
Case Western Reserve University, Cleveland OH 44106-7215 
\\ electronic mail: denise@smaug.astr.cwru.edu}



\begin{abstract}

We show that there is a subsystem of the M31 globular clusters with
{\bf thin disk} kinematics. These clusters span the entire metallicity
range of the M31 globular cluster system, in contrast to the (thick)
disk globulars in the Milky Way which are predominantly
metal-rich. Disk globular clusters are found across the entire disk of
M31 and form $\sim$40\% of the clusters projected on its disk.  The
existence of such a disk system suggests that there was a relatively
large thin disk in place very early in M31's history.  Accurate
measures of the ages of these clusters will constrain the epoch of
disk formation in M31. There is currently no strong evidence for
differences in age between Milky Way and M31 globulars. While age
differences are subtle for old populations, it is unlikely that disk
clusters with \fe\ around --2.0 were formed after significant star
formation began in the galaxy, as the proto-cluster gas would be
enriched by supernova ejecta. Thus it is likely that M31 had a rather
large disk in place at early epochs.

The very existence of such a cold disk means that M31 has suffered no
mergers with an object of 10\% or more of the disk mass since the
clusters were formed. This makes Brown et al (2003)'s suggestion that
M31 could have suffered an equal-mass merger 6-8 Gyr ago less viable.

\end{abstract}


\keywords{clusters: globular}

%% chosen the first three characters of the first author's name plu s

\section{Introduction}

When did present-day galaxy disks start to form?  Despite much
progress in the study of high-redshift galaxies, with almost 1000
galaxies now known with $z>3$, \citep{shapley03}, our observational
constraints on the formation of disk galaxies are much poorer.  There
are few confirmed disk galaxies with redshifts much greater than 1 (a
lookback time of approximately 8 Gyr).  The lack of known
high-redshift disk galaxies is due to two reasons. First, the lower
star formation rate typical of disks leads to a low surface brightness
compared to starbursting galaxies such as the Lyman Break galaxies
\citep{steidel,adelberger}.  This, combined with the $(1+z)^{-4}$
cosmological surface brightness dimming, makes them challenging
observational targets. Second, we need resolved kinematics to confirm
the rotational signature of a disk and linewidths to measure its
luminosity via Tully-Fisher.  Even studies with the Keck telescope
have not succeeded in pushing much beyond z=1, although deep IR
imaging and spectroscopy are now identifying a few more
\citep{vandokkum,labbe,erb}.  Although damped Ly-alpha absorbers in
QSO spectra were originally thought to be young disk galaxies,
evidence is now building that many of these gas-rich objects are in
fact dwarfs \citep{rao}. Measurements of the age of disk stars in the
Milky Way (see below) show significant numbers with ages of 10 Gyr or
more, suggesting that the epoch of disk formation has not yet been
seen at high redshift.

Thus constraints on the epoch and process of disk formation from more local
stellar populations are important.  In the Milky Way, the most accurate
measurement of ages comes from white dwarf cooling and isochrone fits to star
clusters. While most estimates of the age of the local galactic disk from
white dwarfs are close to 10 Gyr \citep{leggett,knox}, uncertainties due to
theoretical models may be as much as 3 Gyr \citep{moroni}. The ``gold
standard" of age determinations is provided by star clusters. The oldest
clusters in the Milky Way (and so the ones that are likely to give
interesting constraints on its formation) are its globular clusters.

In the Milky Way, the dominant stellar populations are the disk, bulge/bar,
and halo. The majority of globular clusters belong to the metal-poor,
pressure-supported halo.  There are no Milky Way globular clusters known to be
associated with its thin disk: the only thin disk clusters are the open
clusters, which are typically several orders of magnitude less luminous than
the globular clusters and usually much younger \citep{eileenaraa}. 

Early work on the Milky Way globulars \citep{baade,kinman} found that
there was a strong correlation between kinematics and metallicity,
with the more numerous metal-poor clusters associated with the
halo. The population membership of the metal-rich clusters is much
less clear.  They have been associated variously with the bulge
\citep{frenkwhite82,dante,cote99}, the thick disk
\citep{zinn85,taft1989} and the bar \citep{cote99,dinescu03}.  Bar and thick
disk clusters probably formed originally in a thin disk and were then
heated into bar or thick disk via various dynamical processes
\citep[eg][]{quinn86,jenkins}. Bulge clusters were likely formed
separately from the disk. The
correct population for the old, metal-rich clusters remains
contentious in many cases because their distance uncertainties
propagate to large errors in kinematical parameters. 

What can we learn about disk formation and evolution from the old
stellar populations of the nearest large spiral, M31? Early
spectroscopic studies of its globular clusters \citep{hbrodie} found
that they had similar kinematic and chemical properties to the Milky
Way system: the metal-richer clusters showing some degree of
rotational support and metal-poor clusters showing little rotation and
high velocity dispersion.  However, as the quality and quantity of
data available on M31 has increased, it has become increasingly clear
that its old stellar populations are {\bf not} simply ``like the Milky
Way's, only more so'' \citep{huchra}. While the Milky Way's halo
dominates more than a few kpc from its disk, leading the practice of
calling similar positions in M31 ``halo'' regions, M31's bulge is much
more luminous and extended than the Milky Way's, with an $R^{1/4}$
profile which extends smoothly from 200 pc to 20 kpc on the minor axis
\citep{pvdb}.  Color-magnitude diagrams of fields tens of kpc from the
major axis have found that stars have higher metallicities and a
larger range in ages than seen in the Milky Way halo
\citep{jeremy,durrell94,rich96,durrell01,tombrown}.  \citet{jeremy}
and \citet{kcf} suggested that the M31 bulge might be the dominant
component even in regions far from the disk.

Kinematical measurements add important information about formation
processes. Examples are the cold, rotationally supported disks
which form by slow dissipative collapse, the moderately
rotating bulges and thick disks whose formation processes combine some
memory of rotation with some heating processes, and the non-rotating
kinematics of the Milky Way's halo which likely formed via
accretion of very small satellites into the galaxy potential. Do the
kinematics of old populations in M31 bear out the suggestion that its
bulge dominates the non-disk regions even at distances of tens of kpc
from its center? Absorption-line spectroscopy of the bright inner
regions of M31's bulge has shown that it has moderate rotational
support \citep{mcelroy,korm88}. Away from these bright regions,
planetary nebulae (PNe) and globular clusters are the tracers of
choice.  \citet{denise}, in their study of the kinematics of PNe out
to distances of 20 kpc from M31's center, found that almost all their
PNe had significant rotational support, with only a few showing the
pressure-supported kinematics typical of the halo. This adds more
weight to the hypothesis that the bulge dominates far from the plane.

Globular clusters generally trace an older age range than PNe, which
are produced by populations more than $\sim$1 Gyr old. \citet{perrett}
showed that the relatively small sample size and large velocity errors
on the original velocities of M31 globulars obtained by
\citet{hbrodie} led to the misconception that many M31 globulars
belonged to a pressure-supported halo like the Milky Way's.
\citet{perrett} used their higher-precision data (which had velocity
measurement errors of only 12 km/s) to show that the rotational
support of the whole system had been under-estimated. The entire M31
globular cluster system has $V/\sigma \sim 1$, with only a slight
correlation between kinematics and metallicity. These kinematics, with
their moderate rotational support, are closer to the bulge than to a
non-rotating halo.  Thus, while M31 does possess some stars in a
Milky-Way-like halo, this halo does not dominate the regions away from
the disk.  The moderately-rotating, $R^{1/4}$ bulge is the major
kinematical component in both PNe (which represent stars with ages
greater than 1 Gyr) and the globular clusters, which represent an even
older population.

What of M31's disk? Is it, too, fundamentally different from the Milky
Way's?  We will show below that M31's thin disk contains a system of
(presumably ancient) globular clusters. An even more profound
difference would be provided by the suggestion of \citet{tombrown}
that there was a merger between M31 and an equal-mass companion as
recently as 6--8 Gyr ago, which is one of their suggestions to explain
the young stars they detect far out on the minor axis in their very
deep HST ACS images.  A merger with such a massive companion would
have destroyed M31's existing disk 
\citep[eg][]{barneshernquist}. The disk we see today would have formed
after the merger.  Since any globular clusters in a thin disk would be
heated into a spheroid by an equal-mass merger, our result leads to
further constraints on M31's merger history.

In Section \ref{s:obsns} we will describe our observational evidence
for the system of thin disk clusters and the kinematic models we use
to interpret their motions projected onto the line of sight. Section
\ref{s:properties} discusses the properties of the thin disk cluster
population and shows that they are found across the entire disk of M31
and their metallicity distribution is quite similar to the overall
cluster distribution.  Section \ref{s:discussion} compares M31's
globular cluster system to the Milky Way's and other disk galaxies,
and concludes with a short discussion of the implications of our
result for disk formation and evolution theories.

\section{Observations}
\label{s:obsns}

\begin{figure}
\includegraphics[scale=0.7]{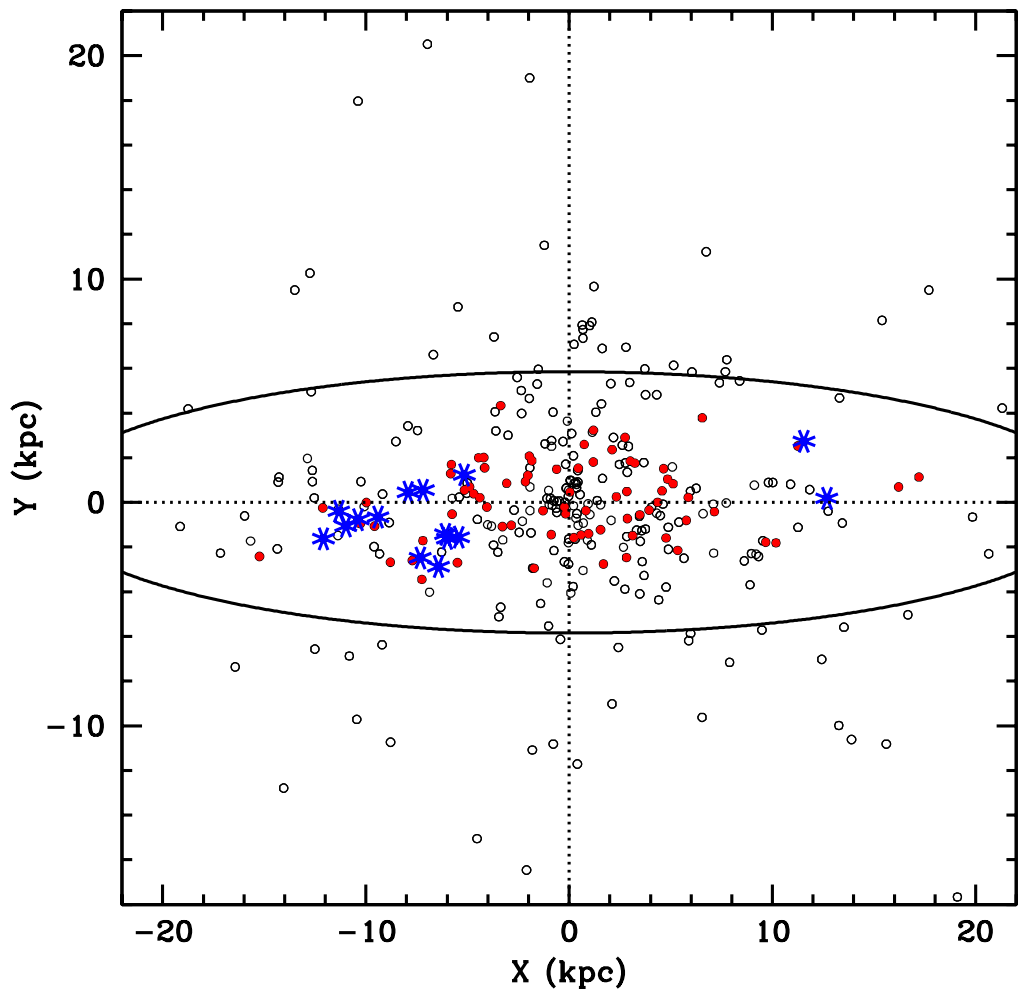}        

\caption{Positions of the M31 globular clusters in the \citet{perrett}
``best current sample''. X is the distance measured parallel to the
major axis, and Y the distance parallel to the minor axis.  Clusters
belonging to the rotating subsample are shown with red, filled symbols
(for those with \fe$<-2.0$) and blue, starred symbols (for those with
\fe$>-2.0$). The edge of the disk at R=26 kpc \citep{wk88} is shown
by a solid line.
\label{xyplot}} 

\end{figure}

We use the ``best current sample'' of \citet{perrett}. This sample is
a compilation of 321 velocities and 301 metallicities for M31 globular
clusters. 225 (65\%) of the velocities have errors less than or equal
to 12 km/s, making this dataset particularly well-suited for
kinematical investigations. Because \citet{perrett} were prevented
from observing fields far from the major axis by poor weather, the
majority of the ``best current sample'' lies within a projected
distance of 5 kpc of the major axis, as can be seen in Figure
\ref{xyplot}. Since the edge of M31's disk projects to a minor axis
distance of $\sim$5 kpc, this means that these clusters with
high-quality velocities are in the best region to detect any disk
globular cluster system.

\citet{perrett} showed that the majority of the globular cluster
sample had significant rotational support. Two of M31's known stellar
populations (disk and bulge) have significant rotation: we will first
compare the kinematics of the globular clusters with these
populations.  We also explore the possibility that the M31 globular
clusters have thick disk kinematics.  The existence of a kinematic
thick disk in M31 is currently unclear, largely due to the lack of
suitable kinematic data at low surface-brightness levels. We consider
models with thick disk kinematics for two reasons. First, \citet{vdks1,vdks2}
found, in their study of edge-on disk galaxies, that galaxies with
large bulges had thick disks, so we might expect M31 to have a thick
disk if this is a general trend. Also, some globular clusters in
the Milky Way have thick disk kinematics, so it is worth checking
whether the M31 clusters do so too.

Plotting the velocity versus the distance along the major axis (X) for
strips of different distance from the major axis (Y) provides a
sensitive diagnostic of disk kinematics \citep{denise}. This is
because, for a population dominated by rotation, we view objects near
the major axis at the ``tangent point'' where their rotational
velocity is projected directly onto the line of sight. The velocities
of objects in strips offset in Y show a tilted line in the X
vs. velocity diagram. This is due to the decreasing amount of
rotational velocity which lies along the line of sight as we move
closer to the minor axis. Once one reaches the inner, solid-body
region of the rotation curve, this too produces a tilted line in the X
vs. velocity diagram.

\begin{figure}
\includegraphics[scale=0.8]{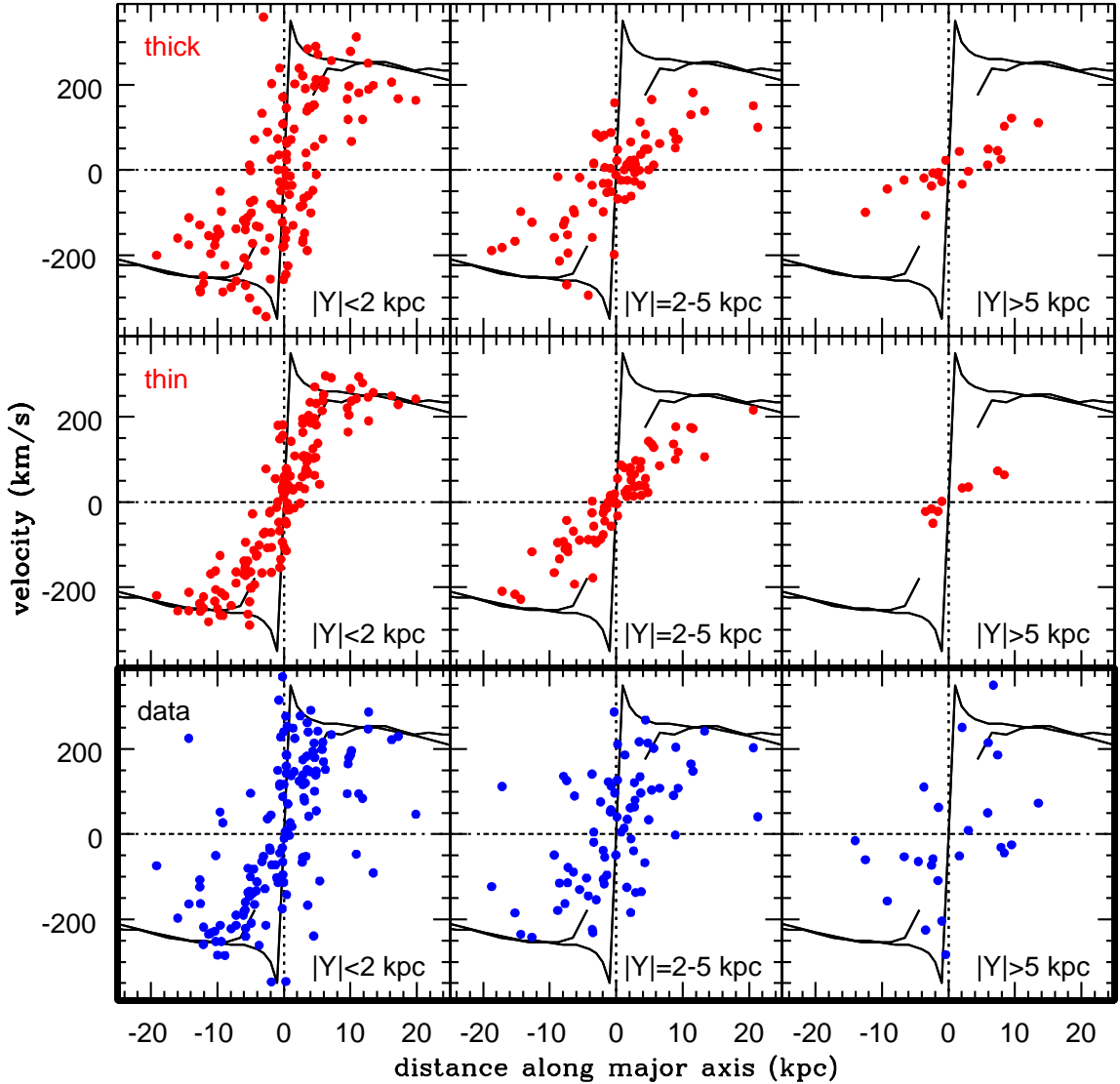}

\caption{Plot of velocity with respect to M31 against distance along
the major axis X for different slices of the distance Y from the major
axis. The bottom three panels show clusters in the best current sample
with velocity errors less than 20 km/s. For comparison, the rotation
curves of \citet{kent, braun} are also shown. The top panels show the
prediction of our thin (middle) and thick (top) disk models for
M31. It can be seen that many of the clusters with $|Y|$ less than 2
kpc have kinematics similar to the thin disk model, while the clusters
with $|Y|$ greater than 2 kpc show a higher velocity dispersion. Note
also that the thin disk model predicts very few clusters with $|Y|>5$
kpc because it assumes that the stellar disk ends at a radius of 26
kpc, which projects to 5.5 kpc on the minor axis \citep{wk88}.
\label{ycuts}} 

\end{figure}

We show the kinematics of the M31 clusters in the lower panels of
Figure \ref{ycuts} for different ranges of Y. The rotation curves of
\citet{kent, braun} are also shown. It is clear that there
is a difference in the kinematics of the clusters close to, and
further away from, the major axis. A distinct line with little
velocity dispersion can be seen in the velocity vs. X plot for $|Y|<2$
kpc: the signature of a kinematically cold, rapidly rotating disk. 

This is particularly noticeable in the lower left panel of Figure
\ref{ycuts} in two regions: the concentration of clusters close to the
circular velocity around X=--10 kpc, and the line stretching from
X=--8 kpc, velocity = --225 km/s, to X=--4 kpc, velocity = --100 km/s.
This turnover is not what we would expect if the disk rotation curve
remains flat until R = 4 kpc or less, as \citet{braun} claims.

\subsection{Kinematical Models of Disk and Bulge}
\label{s:models}

We have made representative kinematical models for both thin and thick
disks and M31's bulge and projected them onto the line of sight. These
models show how kinematics change as the population's mean rotational
velocity decreases and its velocity dispersion increases.  We first
discuss a model for a completely cold disk with no velocity
dispersion, and then extend it to a more realistic model.

\subsubsection{Cold disk model}

Our cold disk model has zero thickness. In this case, the observed
positions on the sky (X and Y) of the clusters uniquely determine
their position in the disk. Because the cold disk has zero velocity
dispersion, the rotation curve uniquely determines the expected
velocity for that position.  There is a surprisingly little agreement
in the literature on the form of M31's rotation curve, and we discuss
the different estimates in Section \ref{s:rotcurve} below, and use the
kinematics of the disk globular clusters themselves to further
constrain its properties. In our cold disk model we use a simple
parameterization which is flat for $|R|>6.5$ kpc and then falls
linearly to zero at X=0.

\begin{figure}
\includegraphics[scale=0.7]{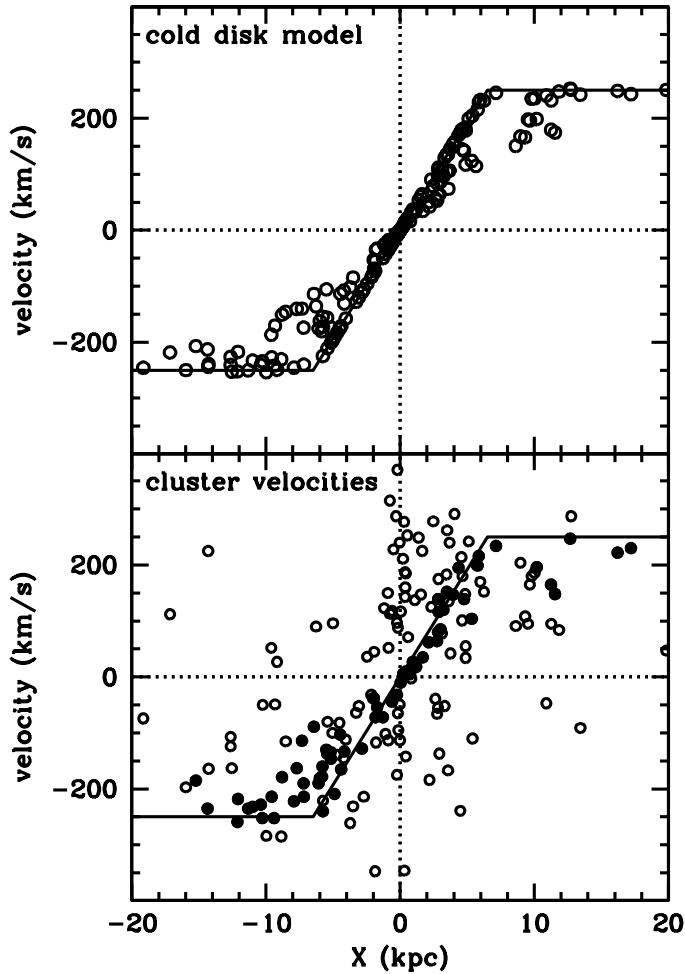}

\caption{Plot of predictions of a completely cold disk model for
clusters with velocity errors less than 20 km/s and $|Y|<2$. Actual
velocities are shown in the lower panel, while
predicted velocities from the cold disk model at that X,Y position are
shown with open symbols in the upper panel. Clusters with velocity
within 30 km/s of the model value at that point are shown with filled
circles in the lower panel. The \citet{kent} rotation
curve is shown for comparison in each panel. Outside the solid-body
region, the effects of
projection can be seen clearly in the model points: even with no
velocity dispersion, there are a range of velocities at a given X
position, with the points with larger $|Y|$ values having smaller
velocities.
\label{centerline1}} 
\end{figure}

We show in Figure \ref{centerline1} the velocities for the cold disk
for each cluster position in the $|Y|<2$ sample, compared to the
actual velocity of the cluster.  Outside the solid-body region, at a
given X value larger Y means that less of the circular velocity is
projected on the line of sight, so the cold disk model velocity is
smaller.  It can be seen that some clusters have velocities very close
to the values expected from a cold disk, while others are hundreds of
km/s away. Clusters with velocity within 30 km/s of
the expected cold disk velocity are shown with solid circles.

\subsubsection{Realistic disk models}
\label{s:realistic}

In fact, real disks have non-zero thickness and velocity dispersion.
For each disk model we specify a spatial distribution, a mean
rotational velocity and a velocity ellipsoid. Details are given in
Table \ref{modelparams}.  The luminosity density of thin and thick
disk models drops exponentially with (cylindrical) R and with height
above the plane z. We have chosen the maximum radius of the thin disk
using the surface photometry of \citet{wk88} and, somewhat
arbitrarily, made the thick disk extend to larger radius.

\begin{deluxetable}{lccccc}
\tablewidth{0pt}
\tablecaption{Parameters for Kinematical Models}
\tablehead{
\colhead{Model} & \colhead{$h_R$ (kpc)} & \colhead{$h_z$ (kpc)} &
\colhead{$R_{max}$ (kpc)} & \colhead{$R_{eff}$ (kpc)} &
\colhead{$V_{rot}$ (km/s)}
}
\startdata

Thin Disk &  5.3 & 0.3  & 26 & \nodata & 250\\
Thick Disk & 5.3 & 1.0 & 40 & \nodata & 210 \\
Bulge      & \nodata & \nodata & \nodata & 2.4 & 110 \\

\enddata

\label{modelparams}
\end{deluxetable}

We know the position on the plane of the sky (X,Y) of each cluster,
but do not know its distance along the line of sight. We make a single
realization of the model by drawing this line of sight distance at
random from a probability distribution determined by the luminosity
distribution of the component at this X,Y position. Thus for the thin
disk model, most clusters will be assigned line of sight distances
within a few hundred kpc of the disk plane, while the thick
disk model will produce positions which cover a larger range of
distances from the plane.

Once we have simulated a 3-dimensional position in M31 for the
cluster, we assign it a velocity using the corresponding kinematical
model for that population. 
Ideally, we would be guided by absorption-line studies of the stellar
populations in M31 in making the best kinematical models. However,
obtaining good absorption-line kinematics for stellar populations in
M31 is difficult because the galaxy covers such a large region of the
sky that good sky-subtraction is very hard. This is illustrated by the
large variation of velocity values in different but symmetrical slit
positions of \citet{mcelroy}.  So we use a combination of M31 data,
kinematics from similar external galaxies and kinematics from the
Milky Way. We do not intend these models to be definitive, but merely
to illustrate the different kinematical signatures of each population.

The kinematics for the thin and thick disk models are based on the
extensive study of the kinematics of disk galaxies of \citet{bottema}
and the M31 disk rotation curve \citep{kent,braun}.  The run of
velocity dispersion with radius in the disk is given by
\[ \sigma_R(R=R) = \sigma_R(R=0)\, e^{(-R/2h_r)} \] 

with \[ \sigma_R(R=h_R) = 0.31 V_{circ} \] 

The exponential decrease of the velocity dispersion with $R/2h_r$ is
needed to give the constant disk scale height with radius that is
observed in many disk galaxies \citep{vdks1,mbh5907}.

The disk circular velocity $V_{circ}$ is fixed to 250 km/s
\citep{kent,braun} for $|X|>6.5$ kpc, falling linearly to 0 at the
disk center. The disk velocity ellipsoid is given by the simple relation
$(\sigma_R,\sigma_{\phi},\sigma_z)=(\sigma_R,\sigma_{R}/\sqrt{2},
\sigma_{R}/2)$ \citep[the epicycle approximation, see][]{binney}.

Less is known about the kinematics of thick disks because of their low
surface brightness, so we use the solar neighborhood values of the
Milky Way thick disk kinematics as a guide: a decrease of 40 km/s to
the rotational velocity \citep{cll} and a velocity dispersion a factor
of 2 higher, with the same variation with cylindrical radius R as the
thin disk.

Once we have simulated the kinematics of each cluster using these
distributions, we project the velocity vector on the line of sight to
produce one realization of what we would observe from an object
belonging to each of the three populations at that X and Y.

\subsubsection{Bulge model}

The bulge luminosity distribution has an $R^{1/4}$ profile with major
axis effective radius $R_{eff}$=2.4 kpc and b/a=0.55
\citep{wk88,pvdb}. Note that \citet{pvdb} show that M31's minor axis
luminosity profile follows this $R^{1/4}$ law for a remarkably large
distance: from R=200 pc to 20 kpc. This suggests the possibility that
M31's bulge may dominate for tens of kpc away from the major axis
\citep{jeremy,kcf}, and so a significant number of globular clusters
might be associated with the bulge.

We considered the observations of bulge kinematics of
\citet{mcelroy,korm88}, and originally constructed a simple isotropic
oblate rotator model with mean rotation $V_{rot}$=80 km/s, decreasing
in concert with the rotation curve inside $|X|$=6.5 kpc, and velocity
dispersion falling linearly with radius from 150 km/s at the center to
100 km/s beyond a radius of 2 kpc. However, both the globular cluster
kinematics far from the plane and the kinematics of planetary nebulae
in M31 \citep{denise} show more rotational support than this.
Also, the kinematics of low-luminosity elliptical galaxies
formed by unequal-mass mergers show an increase of rotational velocity
with increasing radius \citep{naab}.
Thus we have increased the mean rotational velocity to $V_{rot}$=110 km/s.

\subsection{Cluster Kinematics}

\subsubsection{Clusters close to the major axis}

Figure \ref{ycuts} compares the kinematics of the cluster sample to
realizations of the thin (middle panel) and thick (upper panel)
disk. It can be seen that the kinematics of many of the clusters with
$|Y|<2$ kpc more closely resemble the thin disk model, although there
is also a contribution from a hotter population with less rotation. In
particular, note the concentration of clusters in a tight line
stretching from X=--10 kpc, velocity = --250 km/s, to $X\sim 8$ kpc,
velocity = 250 km/s. While the thick disk model does show some
clumping of objects in the X vs. velocity diagram, it is not as
organized as the actual cluster data. Also, there are significantly
more objects in both forbidden quadrants for the thick disk model,
because its larger velocity dispersion allows them to appear there.

In fact, the linear feature in the X--velocity diagram for the
clusters with $|Y|<2$ kpc looks {\bf cooler} than our thin disk model,
showing a tighter relation between X and velocity. We will return to
this below in Section \ref{s:diskinematics}.

The model predictions for $|Y|>2$ kpc show the characteristic gently
sloping line in the X vs. velocity plot which is caused by the fact
that objects close to the minor axis have most of their rotational
velocity projected away from the line of sight. While there is some
resemblance between the thick disk model and the globular clusters for
intermediate values of $|Y|$ (2 to 5 kpc), it is clear that the
objects with $|Y|>5$ kpc are not drawn from a disk population.

\subsubsection{Clusters outside the disk}

\begin{figure}
\includegraphics[scale=0.65,angle=270]{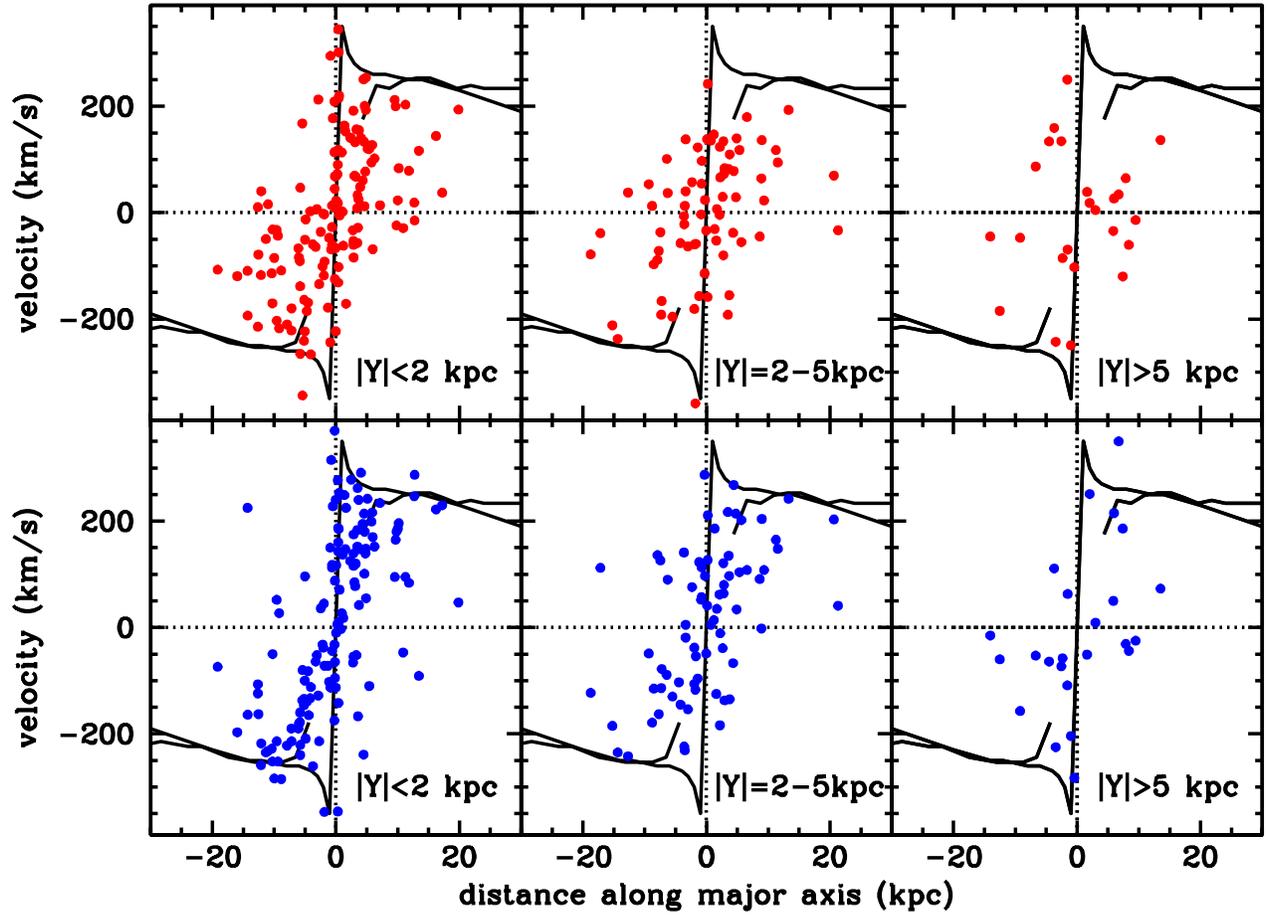}        

\caption{Comparison of the kinematics of the globular clusters (lower
  panels, as in Figure \ref{ycuts}) with
  one realization of the oblate isotropic rotator model for M31's
  bulge (upper panels). 
\label{bulge}} 
\end{figure}

Almost all clusters with $|Y|>5$ kpc are outside the region occupied
by M31's disk, so these clusters are similar to the ``halo'' clusters
studied by previous workers. However, \citet{denise} have shown that
the PNe  in this region have moderate rotational support, which is 
more similar to the kinematics of M31's bulge than to a non-rotating,
pressure-supported halo. It is also clear that the cluster kinematics,
even this far from the major axis, also show significant rotational
support. Thus the bulge is the most likely population for them to
belong to.
Figure \ref{bulge} compares the kinematics of the
globular clusters with one realization of
the bulge model described above. It is interesting to note that even a
rotational velocity of 115 km/s does not lead to a large difference in
the number of objects in the rotating and forbidden quadrants of the
X-velocity plot: the large velocity dispersion of the bulge dominates
its kinematics here. The intermediate-Y panels ($|Y|=2-5$ kpc) show as
much resemblance to the thick disk model as to the bulge one; but it
is clear that there is a reasonable match between the kinematics of
the $|Y|>5$ clusters and the bulge model.

For simplicity, we conclude that the kinematics of the M31 globulars
can be explained using two kinematical components: a thin, rapidly
rotating disk with a low velocity dispersion, and a higher velocity
dispersion component whose properties resemble that of M31's
bulge. Within 2 kpc projected distance from the major axis, 
roughly half of the globular
clusters belong to the thin disk component.

We note here the significant difference between the M31 and Milky Way
globular cluster kinematics. The Milky Way clusters divide into a
predominantly disk
subsystem with kinematics and spatial distribution like a thick disk
and a halo subsystem with high velocity dispersion and little or no
mean rotation. By contrast, the M31 clusters divide into a group with
{\bf thin} disk kinematics and another with high velocity dispersion
but significant rotational support, whose kinematics resemble M31's
large bulge. While there may be a few clusters which have kinematics
typical of a non-rotating halo like the Milky Way's, they do not
dominate the sample. While this sample is restricted primarily to
objects within 5 kpc of the major axis, we note that \citet{denise}
also find significant rotational support in the kinematics of
planetary nebulae up to 20 kpc from M31's major axis.

Before we discuss the cluster kinematics further, we will
investigate the appearance of asymmetry in Figure \ref{ycuts}. The
left hand side (negative X) shows a clearer  rotation signature  than
the right hand side. 
In the Appendix, we use the HST data compiled by \citet{bh01} to
quantify the incompleteness of the M31 globular cluster samples. We
find that the sample of clusters with good velocity and reddening
measurements becomes incomplete just fainter than the peak of the
luminosity function, and that it is more incomplete for positive than
for negative X. The disk clusters are more likely to be found within
M31's dust lane, and so the positive  X side will have a higher
proportion of bulge clusters. 

\subsection{Kinematical Parameters for Thin Disk Subsystem}

While we concentrate here on the kinematics of the thin disk system,
we need to make sure that the bulge clusters found in the same region
of the galaxy do not bias our answers. We do this by using the
clusters whose projected velocities contain the largest fraction of
$V_\phi$, the velocity component which varies the most between the two
systems.  We have calculated the fraction of $V_R, V_\phi$ and $V_z$
which are projected onto the line of sight under the assumption that
the cluster in question is located exactly in the plane of M31's
disk. This will be quite a good assumption for the thin disk clusters,
and less so for the bulge ones. $V_z$ always projects 22\% onto the
line of sight, while the $V_R$ and $V_\phi$ components vary
considerably depending on the X,Y positions of the cluster. If a
cluster is close to the major axis and more than a few kpc from the
center it has a large proportion of $V_\phi$ projected onto the line
of sight.

While there is a difference in velocity dispersion between disk and
bulge for both $V_R$ and $V_\phi$, the $\phi$ component has the
advantage of a difference in mean velocity as well (beyond 6.5 kpc, the
thin disk has a mean velocity of $\sim$250 km/s while the bulge has a
mean velocity of around 100 km/s \citep{mcelroy,denise}. Thus we have
chosen to use clusters which have 50\% or more of their $V_\phi$
velocity projected into the line of sight when we derive kinematic
parameters.

\subsubsection{Disk rotation curve}
\label{s:rotcurve}

Both the velocity dispersion and the deviation from the circular
velocity give us information about the amount of heating
that the clusters have experienced. In the extreme case of a thick
disk, the heating is caused by an event such as a minor merger, but
even an isolated but realistic disk will show an increase in velocity
dispersion with time caused by gravitational interactions with giant
molecular clouds and spiral arms \citep{spitzer1,spitzer2,jenkins}.
First we consider deviations from the circular velocity. 
Despite M31's closeness to the Milky Way, its rotation curve is
surprisingly poorly understood.  The two published estimates of the
rotation curve from HI data are

\begin{itemize}
\item that of \citet{kent}, based on the HI survey of
  \citet{brinkshane} and some HII region data
\item that of \citet{braun}, which is based on a number of HI sources
  but which Braun states ``depends on the good velocity coverage of
  \citet{brinkshane} in the inner galaxy''
\end{itemize}

The two rotation curves are shown in Figure \ref{ycuts}.
There are surprising differences between the two curves,
which are derived at least partially from the same data. This may be
due to the particular problems of determining a rotation curve from a
highly inclined, warped system like M31.  In the case of HI data, the
inner disk velocity field has HI components from the warped, flared
outer disk superimposed. It is not always clear which components are
located in the inner disk.

The two curves agree for values of R between 10 and 20 kpc. Braun's
curve shows a fall of $\sim$50 km/s beyond 20 kpc, while Kent's
remains flat. But it is the differences in the inner galaxy which are
of particular interest here. Braun's curve rises slowly from R = 10
to 3 kpc, and then shows a sharp peak with amplitude $\sim$375 km/s.
Kent's curve only extends to R$>$4 kpc, but shows a gradual fall
inward of 10 kpc. Kent notes that the HII region velocities of
\citet{deharv1,deharv2} show a similar drop to the HI data on the NE
side, which adds credibility to his rotation curve there.

Recent CO observations of M31 help to resolve this discrepancy. Since
molecular gas is generally more centrally concentrated than atomic
gas, it is likely that a rotation curve constructed from CO data will
be less prone to confusion from the warped outer disk. \citet{loinard}
(see also references therein) have made CO observations sensitive enough to
detect the dim emission from the inner disk, and their data are
summarized in \citet{bermloin}. \citet{loin3} show the good overall
agreement between the CO and the \citet{brinkshane} HI observations,
and comment on the large velocity range seen in the inner few
kpc. \citet{bermloin} fit a triaxial bulge model, which successfully
reproduces many features from the CO data. Their rotation curve has a
peak in the central few kpc, but it is less pronounced than
Braun's. Also, between 3 and 10 kpc their rotation curve shows a
gradual fall, not the monotonic rise of Braun's. Because of the
greater agreement between the rotation curves of \citet{kent} and both
the HII region data of \citet{deharv1,deharv2}  and the CO data of 
\citet{bermloin} we will use the Kent rotation curve as a guide in
what follows.

We can also use the velocities of the disk clusters themselves to
constrain the rotation curve. We find that their kinematics are best
fit by a rotation curve with a slow turnover starting at R=6.5 kpc
rather than one which stays flat or rising until $\sim$2 kpc and then
peaks. The reason for this can be seen in Figures \ref{ycuts} and
\ref{centerline1}. If the rotation curve stayed flat or even rose
inside 10 kpc, we would not see the almost linear feature in velocity
for X between --4 and --6 kpc: we would instead expect to see the
cluster velocities stay close to --250 km/s.

To put this argument on a more quantitative basis, we select
clusters with 50\% or more of their $V_\phi$ velocity projected into
the line of sight, and with $|Y|<3$ (to maximise the contribution from
disk clusters).  We then calculate the expected velocity from a
cluster in a cold disk (velocity dispersion = 0) at that X and Y
position, and calculate the difference between the cluster's true
velocity and this expected velocity. We consider disks with a range of
rotation curves. Each rotation curve is flat with $V_{circ}$ =
250 km/s until it reaches the turnover point Rturn, and then decreases
linearly to zero for R=0. We considered values of Rturn between 3 and
10 kpc and calculated $\chi^2$, the sum of square of the residuals,
for each rotation curve, finding a distinct minimum for a turnover
radius of Rturn=6.5 kpc. We have used this simple rotation curve for
all the kinematic models in this paper.

\subsubsection{Estimating disk kinematics}
\label{s:diskinematics}

At this stage, the relatively small sample size makes it difficult to
produce exact estimates of the kinematics of the disk subsystem. Until
we obtain a larger and more complete sample of velocities of the disk
clusters, we will estimate their kinematics by eye. This graphical
method also allows the clusters which likely do not belong to the disk
to be discounted.

\begin{figure}
\includegraphics[scale=0.7]{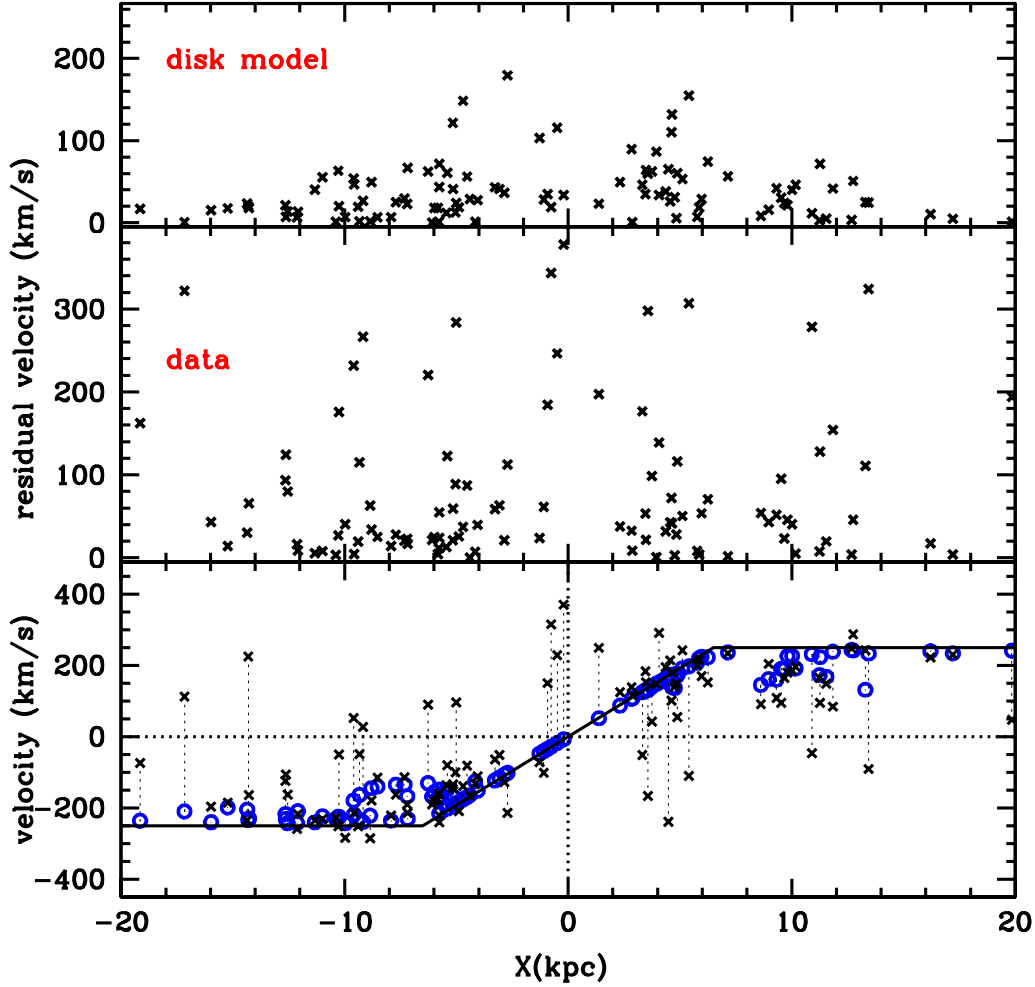}        

\caption{Comparison of velocity values for clusters with $|Y|<5$ kpc which have 50\% or more of
their $V_\phi$ velocity projected into the line of sight with both a
completely cold disk and the more realistic disk model described in
Section \ref{s:realistic}. The bottom panel shows observed velocities
(crosses) joined by dotted lines to the corresponding velocity of a
completely cold disk object at the same X and Y. The middle and top panels
show the absolute value of the difference between real and cold disk
velocity, for both clusters (middle panel) and for the realistic disk
model described in Section \ref{s:realistic}. The real data shown in
the middle panel are well fit by a combination of the disk model of
the upper panel and a hotter component which we associate with the bulge.
\label{velresid}} 
\end{figure}

Figure \ref{velresid} plots the velocity residuals from a cold disk
against X for all clusters with $|Y|<5$ kpc which have 50\% or more of
their $V_\phi$ velocity projected into the line of sight.  The lower
panel shows real velocities (crosses) joined by a dotted line to the
predicted velocity from a cold disk object at that X and Y (open
circles). The length of the dotted line is the residual plotted in the
upper two panels, for both real data (middle panel) and the thin disk
model of Table 1 (top panel).  It can be seen that the velocity
residual panel for the real data is well explained by a combination of
the disk model shown above and a population with higher velocity
dispersion which we have identified with the bulge.

\section{Properties of the Thin Disk Clusters}
\label{s:properties} 

\subsection{Deciding Disk Membership}

When two populations have spatial and kinematic overlaps, as in this
case, it is not possible to allocate a cluster with 100\% accuracy to
a certain population. The problem is accentuated here because there is
some uncertainty in the kinematic parameters of both populations. We
have chosen to measure the difference between the actual cluster
velocity and the velocity of an object belonging to a totally cold
disk at the same X and Y.  This residual will be small for disk
objects and larger, on average, for bulge objects.

Because the velocity dispersion of a real disk with constant scale
height increases towards its center, we normalize this residual using
a measure of velocity spread obtained from 100 realizations of our best fit
disk model for each X and Y. The interquartile range of disk model
velocities gives a non-parametric estimate of the expected spread of
velocities from the model disk, and we divide by it to normalize the
residual appropriately. 

\begin{figure}
\includegraphics[scale=0.5]{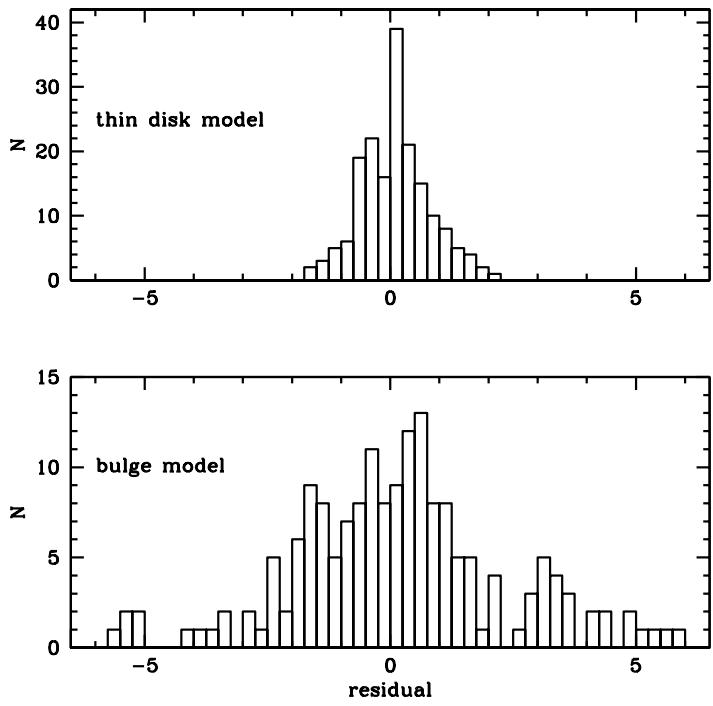}        

\caption{Histogram of residuals calculated as described in the text
  for a model drawn from the disk parameters from Table \ref{modelparams}
  (upper panel) and for a model drawn from the bulge parameters given
  in Table \ref{modelparams}.
\label{fakeresidhist}} 
\end{figure}

Thus for a given cluster, the residual is defined to be:

\[ \frac{V_{actual}-V_{cold\, disk}}{IQR\,\,{\rm of\, disk\, model}}
\]
We use our simulations to describe the distribution of residuals from
both a pure disk and a pure bulge population. In both cases we have
taken a single realization of the kinematic models and calculated the
residuals as described above. Figure \ref{fakeresidhist} shows the
behaviour of these residuals.  It can be seen that the disk objects
cluster around a residual of 0: all but a couple have residuals
between --2 and 2; in fact 83\% have residuals between --1 and 1 and
68\% have residuals between -0.75 and 0.75. By contrast, 31\% (28\%)
of the bulge model objects have residuals between --1 and 1 (--0.75
and 0.75). Thus, classifying objects with residual between --0.75 and
0.75 as disk members gives a $\sim$30\% chance of mis-classification
in both directions (calling a disk object a bulge member or calling a
bulge object a disk member).

We have chosen to plot various quantities against this residual so
that the effect of another choice of cut in the residual value
defining disk membership (either more conservative or more permissive)
can be readily seen.

Table \ref{classifytable} gives the value of this residual for all
clusters projected on the disk with velocity errors less than 20 km/s.
The letters preceding the identification number in the M31 globular             
cluster names refer to the catalogs listed in Table~\ref{tab:targref}.

\begin{deluxetable}{lccrrrrr}

\rotate
\tablecolumns{8}
\tablewidth{0pt}
\tablecaption{Disk classification residuals for clusters projected on
the M31 disk with velocity errors less than 20 km/s. \label{classifytable}}
\tablehead{
\colhead{ID} & \colhead{Residual} & \colhead{X (arcmin)}  &
\colhead{Y (arcmin)} & \colhead{velocity (km/s)} & 
\colhead{velocity error} & \colhead{[Fe/H]} & 
\colhead{[Fe/H] error}  } 

\startdata
                  	 						     
    B1--S39     &      5.67 &  --34.21 &    14.76 &   --174 &   12 &     --0.58 &   0.18 \\
    B4--S50     &     --2.27 &  --11.72 &    25.70 &   --373 &   12 &     --1.26 &   0.59 \\
    B5--S52     &      4.28 &  --42.10 &     1.72 &   --273 &   12 &     --1.18 &   0.17 \\
    B8--S60   &      0.70 &  --15.46 &    19.88 &   --319 &   12 &     --0.41 &   0.38 \\
    B10--S62     &      5.70 &  --16.72 &    18.61 &   --159 &   12 &     --1.77 &   0.14 \\
    B12--S64   &     --0.98 &  --10.79 &    22.98 &   --358 &   12 &     --1.65 &   0.19 \\
    B13--S65      &     --3.49 &   --7.21 &    24.32 &   --409 &   12 &     --1.01 &   0.49 \\
    B15   &     --0.04 &  --26.57 &     7.78 &   --460 &   12 &     --0.35 &   0.96 \\
    B16--S66     &     --2.60 &   --8.99 &    21.35 &   --406 &   12 &     --0.78 &   0.19 \\
    B17--S70      &     --3.31 &  --16.53 &    14.67 &   --524 &   12 &     --0.42 &   0.45 \\
    B18--S71     &     --1.23 &  --40.63 &    --4.15 &   --585 &   12 &     --1.63 &   0.77 \\
    B19--S72     &      4.07 &  --10.73 &    18.25 &   --224 &    2 &     --1.09 &   0.02 \\
    B21--S75      &      0.16 &  --20.45 &     9.18 &   --403 &   12 &     --0.90 &   0.06 \\
    B23--S78     &     --2.21 &  --13.79 &    13.81 &   --454 &    6 &     --0.92 &   0.10 \\
    B25--S84     &      3.16 &  --22.98 &     3.99 &   --204 &   12 &     --1.46 &   0.13 \\
    B26--S86    &      1.98 &   --3.64 &    18.53 &   --243 &   12 &      0.01 &   0.38 \\
    B28--S88   &      0.72 &  --23.64 &     2.54 &   --434 &   12 &     --1.87 &   0.29 \\
    B29--S90   &     --0.32 &  --22.44 &     3.15 &   --509 &   12 &     --0.32 &   0.14 \\
    B30--S91   &      1.62 &  --24.80 &     1.10 &   --380 &   12 &     --0.39 &   0.36 \\
    B31--S92   &      1.08 &  --23.12 &     1.88 &   --400 &   12 &     --1.22 &   0.40 \\
    B33--S95   &      0.41 &  --21.57 &     1.78 &   --439 &   12 &     --1.33 &   0.24 \\
    B34--S96   &     --0.11 &  --26.43 &    --2.40 &   --540 &   12 &     --1.01 &   0.22 \\
    B37   &      0.25 &   --8.99 &     9.52 &   --338 &   12 &     --1.07 &   0.20 \\
    B38--S98    &      2.25 &   --5.45 &    12.03 &   --177 &   12 &     --1.66 &   0.44 \\
    B39--S101   &      1.60 &   --3.99 &    12.72 &   --248 &   12 &     --0.70 &   0.32 \\
    B40--S102   &     --0.40 &  --35.38 &   --11.94 &   --463 &   12 &     --0.98 &   0.48 \\
    B41--S103   &     --0.13 &   --8.44 &     8.56 &   --372 &   12 &     --1.22 &   0.23 \\
    B42--S104   &      0.67 &  --14.11 &     3.93 &   --352 &   12 &     --0.78 &   0.31 \\
    B43--S106   &      0.42 &  --33.58 &   --11.38 &   --414 &   12 &     --2.42 &   0.51 \\
    B45--S108      &     --5.24 &    7.28 &    20.24 &   --425 &    0 &     --1.05 &   0.25 \\
    B47--S111     &     --1.02 &   13.66 &    24.63 &   --291 &   12 &     --1.62 &   0.41 \\
    B48--S110   &      1.27 &   --8.88 &     7.09 &   --255 &   12 &     --0.40 &   0.37 \\
    B49--S112   &     --0.39 &  --27.49 &    --7.41 &   --481 &   12 &     --2.14 &   0.55 \\
    B50--S113     &      6.32 &    6.04 &    18.52 &   --114 &   12 &     --1.42 &   0.37 \\
    B51--S114   &      0.99 &    0.57 &    14.17 &   --259 &   12 &     --1.00 &   0.13 \\
    B53     &      4.68 &   --1.39 &    12.52 &    --13 &   12 &     --0.33 &   0.26 \\
    B54--S115   &      0.47 &  --18.56 &    --0.96 &   --412 &   12 &     --0.45 &   0.17 \\
    B55--S116   &      0.67 &   --9.34 &     5.56 &   --338 &   12 &     --0.23 &   0.07 \\
    B56--S117   &      1.16 &  --20.72 &    --3.48 &   --382 &   12 &     --0.06 &   0.10 \\
    B57--S118   &      0.27 &  --24.94 &    --7.16 &   --437 &   12 &     --2.12 &   0.32 \\
    B58--S119     &      4.25 &  --28.78 &   --10.21 &   --210 &   16 &     --1.45 &   0.24 \\
    B59--S120   &      0.38 &   --9.86 &     4.29 &   --332 &   12 &     --1.36 &   0.52 \\
    B61--S122   &     --0.23 &    5.49 &    14.82 &   --286 &   12 &     --0.73 &   0.28 \\
    B65--S126   &      0.44 &  --33.24 &   --15.81 &   --378 &   12 &     --1.56 &   0.03 \\
    B66--S128   &      0.70 &  --29.50 &   --13.18 &   --389 &   12 &     --2.10 &   0.35 \\
    B69--S132   &     --0.27 &    3.40 &    11.90 &   --295 &   12 &     --1.35 &   0.43 \\
    B72     &      2.89 &    0.97 &     9.56 &    --89 &   12 &     --0.38 &   0.25 \\
    B74--S135      &     --5.61 &   17.26 &    22.11 &   --435 &   12 &     --1.88 &   0.06 \\
    B75--S136   &      1.02 &   --0.79 &     7.85 &   --212 &   12 &     --1.03 &   0.33 \\
    B76--S138     &     --1.13 &  --12.44 &    --1.58 &   --514 &   12 &     --0.72 &   0.06 \\
    B81--S142   &     --0.47 &  --25.22 &   --12.36 &   --430 &   12 &     --1.74 &   0.40 \\
    B82--S144   &      0.66 &  --15.04 &    --4.95 &   --364 &    6 &     --0.80 &   0.18 \\
    B83--S146      &     --4.15 &   19.78 &    22.10 &   --367 &   12 &     --1.27 &   0.35 \\
    B88--S150      &     --5.36 &    9.99 &    13.35 &   --484 &   12 &     --1.81 &   0.06 \\
    B90   &     --0.43 &  --13.07 &    --4.68 &   --428 &   12 &     --1.39 &   0.80 \\
    B91--S151   &     --0.16 &    2.06 &     7.01 &   --290 &   12 &     --1.80 &   0.61 \\
    B93--S155     &     --2.09 &    1.94 &     6.57 &   --442 &   12 &     --1.03 &   0.12 \\
    B94--S156     &     --1.99 &  --17.08 &    --8.75 &   --561 &   12 &     --0.41 &   0.31 \\
    B97--S159   &     --0.11 &    5.44 &     8.29 &   --282 &   12 &     --1.21 &   0.13 \\
    B102      &      0.17 &   12.63 &    13.34 &   --236 &   12 &     --1.57 &   0.10 \\
    B104     &     --0.76 &   --0.66 &     2.91 &   --395 &   10 &   \nodata & \nodata  \\
    B105--S166   &      0.24 &    9.71 &    10.83 &   --238 &   12 &     --1.13 &   0.32 \\
    B109--S170   &     --0.37 &   --5.88 &    --1.70 &   --372 &   12 &     --0.13 &   0.41 \\
    B110--S172   &      1.49 &  --11.28 &    --6.15 &   --264 &   12 &     --1.06 &   0.12 \\
    B116--S178     &     --1.67 &   12.03 &    11.75 &   --339 &   12 &     --0.88 &   0.12 \\
    B117--S176     &     --2.68 &  --16.10 &   --10.23 &   --531 &   12 &     --1.33 &   0.45 \\
    B118     &      1.62 &   --2.25 &     0.32 &    --72 &   10 &  \nodata & \nodata \\
    B119         &      0.07 &    0.19 &     2.11 &   --310 &   12 &     --0.49 &   0.18 \\
    B122--S181      &     --3.30 &   13.40 &    11.50 &   --437 &   12 &     --1.69 &   0.34 \\
    B124         &      2.64 &   --0.93 &     0.04 &     70 &   13 &   \nodata & \nodata \\
    B125--S183      &     --3.41 &   --8.60 &    --6.24 &   --647 &   12 &     --1.52 &   0.08 \\
    B126--S184   &      1.13 &   --2.76 &    --2.04 &   --182 &   14 &     --1.20 &   0.47 \\
    B127--S185     &     --1.02 &   --1.12 &    --0.92 &   --475 &   12 &     --0.80 &   0.14 \\
    B129   &      1.52 &    7.53 &     4.94 &    --75 &   12 &     --1.21 &   0.32 \\
    B130--S188    &      1.97 &   11.34 &     7.80 &    --22 &   12 &     --1.28 &   0.19 \\
    B134--S190   &     --0.38 &   --0.79 &    --2.37 &   --365 &   12 &     --0.64 &   0.08 \\
    B135--S192     &     --2.14 &   12.68 &     8.12 &   --366 &   12 &     --1.62 &   0.04 \\
    B137--S195    &      0.06 &   13.79 &     8.50 &   --215 &   12 &     --1.21 &   0.29 \\
    B140   &     --0.63 &   --4.04 &    --6.61 &   --413 &   12 &     --0.88 &   0.77 \\
    B141--S197   &      0.37 &   14.83 &     8.06 &   --180 &   12 &     --1.59 &   0.21 \\
    B144   &      0.94 &    1.76 &    --2.33 &   --140 &   12 &     --0.64 &   0.21 \\
    B147--S199   &      1.94 &    6.32 &     0.39 &    --51 &    1 &     --0.24 &   0.36 \\  
    B148--S200   &     --0.44 &    3.78 &    --1.69 &   --303 &   12 &     --1.15 &   0.34 \\
    B149--S201   &      1.69 &   16.86 &     8.18 &    --60 &   12 &     --1.35 &   0.25 \\
    B156--S211     &     --1.27 &   --8.27 &   --13.51 &   --417 &   12 &     --1.51 &   0.38 \\
    B158--S213   &      1.94 &   --3.44 &    --9.88 &   --187 &    1 &     --1.02 &   0.02 \\
    B159   &      0.37 &   10.66 &     1.13 &   --175 &   12 &     --1.58 &   0.41 \\
    B160--S214   &     --0.46 &   --7.89 &   --13.51 &   --354 &   12 &     --1.17 &   1.25 \\
    B161--S215     &     --1.02 &   --0.11 &    --7.51 &   --413 &   12 &     --1.25 &   0.35 \\
    B163--S217   &      0.56 &   13.01 &     2.23 &   --161 &    3 &     --0.36 &   0.27 \\
    B164         &      0.07 &    1.05 &    --7.26 &   --294 &   12 &     --0.09 &   0.40 \\
    B166     &      2.88 &    1.37 &    --7.56 &    --23 &   12 &     --1.33 &   0.37 \\
    B167   &      0.44 &    2.69 &    --6.68 &   --229 &   12 &     --0.42 &   0.23 \\
    B170--S221   &      1.54 &  --15.46 &   --21.52 &   --295 &   12 &     --0.54 &   0.24 \\
    B171--S222   &     --0.04 &    4.37 &    --6.42 &   --273 &    2 &     --0.41 &   0.04 \\
    B176--S227      &     --6.65 &  --15.84 &   --23.49 &   --525 &   12 &     --1.60 &   0.10 \\
    B179--S230   &      0.72 &    7.08 &    --5.62 &   --153 &   12 &     --1.10 &   0.02 \\
    B180--S231   &      1.69 &   --1.07 &   --12.18 &   --203 &   12 &     --1.19 &   0.07 \\
    B182--S233     &     --1.30 &   --0.17 &   --12.64 &   --349 &    4 &     --1.24 &   0.12 \\
    B184--S236   &     --0.20 &   22.23 &     4.78 &   --152 &   12 &     --0.37 &   0.40 \\
    B185--S235   &      1.40 &    5.02 &    --8.71 &   --163 &   12 &     --0.76 &   0.08 \\
    B188--S239    &      0.15 &   13.15 &    --3.33 &   --184 &   12 &     --1.51 &   0.17 \\
    B190--S241   &      0.49 &   20.95 &     2.38 &    --86 &   12 &     --1.03 &   0.09 \\
    B193--S244   &      0.65 &   23.45 &     3.84 &    --58 &    2 &     --0.44 &   0.17 \\
    B197--S247   &      1.30 &   18.61 &    --0.98 &     --9 &   12 &     --0.43 &   0.36 \\
    B198--S249   &      0.33 &   19.98 &     0.02 &   --105 &   12 &     --1.13 &   0.30 \\
    B199--S248     &     --2.45 &   --6.45 &   --20.76 &   --396 &   12 &     --1.59 &   0.11 \\
    B200   &     --0.21 &   18.07 &    --1.57 &   --153 &   12 &     --0.91 &   0.61 \\
    B203--S252   &     --0.91 &   21.19 &    --0.36 &   --199 &   12 &     --0.90 &   0.32 \\
    B204--S254     &     --1.61 &   13.00 &    --7.00 &   --355 &   12 &     --0.80 &   0.17 \\
    B205--S256     &     --2.15 &   15.25 &    --5.67 &   --352 &   19 &     --1.34 &   0.13 \\
    B207--S258     &      3.52 &    0.90 &   --17.27 &   --173 &   12 &     --0.81 &   0.59 \\
    B208--S259   &     --0.29 &   14.33 &    --6.83 &   --222 &   12 &     --0.84 &   0.04 \\
    B209--S261      &     --2.99 &   16.39 &    --5.81 &   --467 &   12 &     --1.37 &   0.13 \\
    B210         &      0.11 &    7.74 &   --12.67 &   --265 &   12 &     --1.90 &   0.32 \\
    B213--S264      &     --4.75 &   20.58 &    --2.71 &   --539 &   12 &     --1.02 &   0.11 \\
    B214--S265     &     --1.26 &   17.22 &    --5.47 &   --258 &   12 &     --1.00 &   0.61 \\
    B216--S267   &     --0.10 &   26.90 &     1.02 &    --84 &   12 &     --1.87 &   0.39 \\
    B217--S269    &      2.30 &   16.07 &    --7.97 &    --38 &   12 &     --0.93 &   0.14 \\
    B218--S272   &      0.36 &   12.95 &   --11.32 &   --220 &    1 &     --1.19 &   0.07 \\
    B219--S271      &     --8.64 &   --4.65 &   --25.40 &   --504 &   12 &     --0.73 &   0.53 \\
    B220--S275     &     --1.54 &   22.38 &    --5.10 &   --245 &   12 &     --1.21 &   0.09 \\
    B221--S276      &     --3.46 &   24.78 &    --4.06 &   --410 &   12 &     --1.29 &   0.04 \\
    B222--S277     &     --1.07 &   10.22 &   --16.12 &   --311 &   12 &     --1.11 &   0.37 \\
    B223--S278   &     --0.01 &   26.45 &    --3.68 &   --101 &   12 &     --1.13 &   0.51 \\
    B224--S279   &     --0.08 &   21.89 &    --7.30 &   --161 &    2 &     --1.80 &   0.05 \\
    B225--S280   &      1.43 &   16.50 &   --12.16 &   --165 &    0 &     --0.67 &   0.12 \\
    B231--S285     &     --1.33 &   22.40 &    --9.65 &   --266 &   12 &     --1.49 &   0.41 \\
    B232--S286    &      2.52 &   12.58 &   --17.82 &   --179 &   12 &     --1.83 &   0.14 \\
    B234--S290   &     --0.27 &   24.51 &    --9.85 &   --196 &   12 &     --0.95 &   0.13 \\
    B235--S297   &      1.47 &   25.93 &   --11.48 &    --98 &   12 &     --0.72 &   0.26 \\
    B237--S299     &      4.54 &   21.89 &   --17.38 &    --86 &   12 &     --2.09 &   0.28 \\
    B238--S301     &      5.80 &   20.22 &   --20.02 &    --32 &   12 &     --0.57 &   0.66 \\
    B272--V294   &      0.46 &   21.32 &     6.93 &   --120 &   12 &     --1.25 &   0.16 \\
    B281--S288   &      0.76 &   16.91 &   --15.01 &   --203 &   12 &     --0.87 &   0.52 \\
    B283--S296     &      5.35 &   15.92 &   --18.80 &    --83 &   12 &     --0.06 &   0.20 \\
    B298--S21      &     --4.22 &  --58.24 &    22.73 &   --542 &   12 &     --2.07 &   0.11 \\
    B301--S22     &      7.37 &  --87.83 &    --4.94 &   --374 &   12 &     --1.22 &   0.18 \\
    B303--S26   &      1.84 &  --65.50 &     5.19 &   --464 &   12 &     --2.09 &   0.41 \\
    B305--DAO24   &      1.63 &  --73.25 &    --2.75 &   --497 &   12 &     --1.29 &   0.57 \\
    B306--S29    &      2.46 &  --57.98 &     6.59 &   --424 &   12 &     --0.64 &   0.40 \\
    B307--S30     &      3.02 &  --57.95 &     4.29 &   --407 &   12 &     --0.41 &   0.36 \\
    B311--S33    &      2.15 &  --57.55 &     0.94 &   --463 &   12 &     --1.96 &   0.07 \\
    B312--S35     &      8.19 &  --36.36 &    15.69 &   --164 &   12 &     --1.41 &   0.08 \\
    B313--S36   &      0.93 &  --39.11 &    12.51 &   --415 &   12 &     --1.09 &   0.10 \\
    B314--S37   &      0.51 &  --69.88 &   --11.09 &   --485 &   12 &     --1.61 &   0.32 \\
    B315--S38   &     --0.18 &  --55.62 &    --1.13 &   --559 &   12 &     --1.88 &   0.52 \\
    B316--S40     &      2.86 &  --47.00 &     4.29 &   --350 &   12 &     --1.47 &   0.23 \\
    B319--S44     &      0.21 &  --51.99 &    --1.81 &   --535 &   12 &     --2.27 &   0.47 \\
    B321--S46   &     --0.05 &  --55.47 &    --7.46 &   --518 &   12 &     --2.39 &   0.41 \\
    B327--S53   &      0.25 &  --47.65 &    --3.51 &   --528 &   12 &     --2.33 &   0.49 \\
    B335      &      0.19 &  --43.95 &    --4.87 &   --514 &   12 &     --1.05 &   0.26 \\
    B338--S76     &      4.85 &  --44.03 &    --9.09 &   --248 &   12 &     --1.34 &   0.08 \\
    B341--S81     &      2.79 &  --42.79 &   --10.58 &   --349 &   12 &     --1.17 &   0.05 \\
    B342--S94   &     --0.66 &  --40.35 &   --12.27 &   --479 &   12 &     --1.62 &   0.02 \\
    B355--S193     &      5.25 &   33.93 &    24.56 &   --114 &   12 &     --1.62 &   0.43 \\
    B356--S206   &      0.43 &   30.06 &    17.38 &   --192 &   12 &     --1.46 &   0.28 \\
    B365--S284     &      4.77 &   61.01 &    21.44 &    --58 &   12 &     --1.35 &   0.14 \\
    B366--S291   &     --0.32 &   51.60 &    11.57 &   --135 &   12 &     --1.79 &   0.05 \\
    B367--S292   &     --0.58 &   52.98 &    12.55 &   --152 &   12 &     --2.32 &   0.53 \\
    B370--S300      &     --6.13 &   49.98 &     3.71 &   --347 &   12 &     --1.80 &   0.02 \\
    B372--S304      &     --3.22 &   54.31 &     2.64 &   --216 &   12 &     --1.42 &   0.17 \\
    B373--S305     &     --1.46 &   43.72 &    --7.86 &   --205 &   18 &     --0.50 &   0.22 \\
    B374--S306   &      0.88 &   41.14 &   --10.50 &    --96 &   12 &     --1.90 &   0.67 \\
    B375--S307     &     --1.24 &   39.53 &   --12.03 &   --209 &   12 &     --1.23 &   0.22 \\
    B378--S311      &     --3.29 &   51.68 &    --5.12 &   --205 &   12 &     --1.64 &   0.26 \\
    B380--S313   &      1.14 &   58.47 &    --1.82 &    --13 &   12 &     --2.31 &   0.45 \\
    B382--S317      &     --3.18 &   40.83 &   --16.92 &   --302 &   12 &     --1.52 &   0.27 \\
    B386--S322      &     --7.57 &   61.66 &    --4.23 &   --391 &   12 &     --1.62 &   0.14 \\
    B400--S343      &     --8.47 &   91.06 &    --3.00 &   --253 &   12 &     --2.01 &   0.21 \\
    B443   &     --0.08 &  --50.41 &    --4.85 &   --532 &   12 &     --2.37 &   0.46 \\
    B448   &     --0.16 &  --43.13 &    --3.01 &   --552 &   12 &     --2.16 &   0.19 \\
    B451     &      0.20 &  --32.95 &     2.50 &   --514 &   12 &     --2.13 &   0.43 \\
    B453   &      0.47 &  --23.69 &     5.64 &   --446 &   12 &     --2.09 &   0.53 \\
    B458   &     --0.83 &  --26.44 &    --6.37 &   --521 &   12 &     --1.18 &   0.67 \\
    B467--S202      &     --6.27 &   38.49 &    24.93 &   --344 &   12 &     --2.49 &   0.47 \\
    B472   &      0.78 &   15.87 &    --2.83 &   --117 &   12 &     --1.45 &   0.02 \\
    B475     &     --1.02 &   44.98 &     4.10 &   --120 &   12 &     --2.00 &   0.14 \\
    B480   &     --0.50 &   44.36 &    --8.18 &   --135 &   12 &     --1.86 &   0.66 \\
    B483   &     --0.09 &   58.17 &     0.84 &    --53 &   12 &     --2.96 &   0.35 \\
    B484--S310    &      0.06 &   46.69 &    --8.31 &   --104 &   12 &     --1.95 &   0.59 \\

    BA11   &     --0.97 &   94.70 &   --10.54 &    --97 &   12 &     --1.14 &   0.61 \\

    BoD195   &     --0.45 &  --47.14 &    --4.38 &   --552 &   12 &     --1.64 &   0.19 \\
    BoD289   &     --0.73 &   74.37 &     3.18 &    --78 &   12 &     --1.71 &   0.63 \\
    BoD292   &     --0.27 &   78.96 &     5.21 &    --70 &   12 &     --0.47 &   0.54 \\

    DAO23      &     13.12 &  --65.65 &     4.16 &    --75 &   12 &     --0.43 &   0.13 \\
    DAO25      &     12.90 &  --78.71 &   --10.42 &   --188 &   12 &     --1.96 &   0.97 \\
    DAO30   &     --0.77 &  --65.86 &    --9.56 &   --535 &   12 &     --0.65 &   0.34 \\
    DAO36   &      0.51 &  --36.33 &     2.07 &   --522 &   12 &     --2.16 &   0.32 \\
    DAO39   &     --0.10 &  --26.73 &     5.92 &   --478 &   12 &     --1.22 &   0.41 \\
    DAO41   &     --0.46 &  --19.27 &     9.21 &   --445 &   12 &     --1.14 &   0.30 \\
    DAO47   &     --0.35 &  --33.01 &    --7.85 &   --490 &   12 &     --1.13 &   0.57 \\
    DAO48   &     --0.07 &  --27.91 &    --6.55 &   --490 &   12 &     --2.01 &   0.99 \\
    DAO58   &      0.79 &   13.13 &     6.19 &   --125 &   12 &     --0.87 &   0.07 \\
    DAO65   &     --0.81 &   27.37 &     2.32 &   --130 &   12 &     --1.80 &   0.36 \\
    DAO66     &     --1.51 &   28.68 &     2.88 &   --148 &   12 &     --1.82 &   0.26 \\
    DAO70   &     --0.08 &   32.78 &    --1.90 &    --66 &   12 &      0.33 &   0.36 \\
    DAO73     &     --1.25 &   45.98 &     4.08 &   --114 &   12 &     --1.99 &   0.19 \\
    DAO84     &     --1.19 &   42.76 &   --11.08 &   --192 &   12 &     --1.79 &   0.72 \\

    NB16   &      1.35 &    1.97 &     4.19 &   --115 &   15 &     --1.36 &   0.12 \\
    NB20   &     --0.86 &   --4.99 &     0.80 &   --402 &   12 &     --0.80 &   0.23 \\
    NB33   &      0.94 &    0.21 &     4.52 &   --183 &   12 &      0.04 &   0.38 \\
    NB61         &     --2.69 &    1.42 &     4.12 &   --646 &   10 &  \nodata  & \nodata \\
    NB67   &      1.66 &    1.70 &     3.74 &   --113 &   17 &     --1.43 &   0.13 \\
    NB68   &      0.93 &    1.70 &     2.95 &   --157 &   11 &     --0.76 &   0.33 \\
    NB74   &      1.48 &   --0.18 &     1.24 &    --60 &   12 &     --0.02 &   0.43 \\
    NB81    &      2.46 &   --3.48 &     0.35 &     15 &   11 &     --0.75 &   0.33 \\
    NB83   &      1.30 &   --4.23 &     0.84 &   --150 &   14 &     --1.26 &   0.16 \\
    NB87   &      1.59 &    2.47 &     1.61 &    --47 &   10 &      0.26 &   0.41 \\
    NB89   &     --0.18 &   --1.09 &    --0.95 &   --332 &    6 &     --0.53 &   0.57 \\
    NB91   &      1.13 &   --2.93 &    --1.19 &   --187 &   10 &     --0.71 &   0.33 \\

    S47   &     --0.62 &  --45.78 &    --0.01 &   --584 &   12 &     --1.19 &   0.29 \\
    S245   &      0.29 &   15.96 &    --2.54 &   --148 &   12 &     --0.31 &   0.16 \\

    V31          &      0.02 &  --19.06 &     7.11 &   --433 &   12 &     --1.59 &   0.06 \\
    V216         &      0.16 &  --20.17 &     0.96 &   --465 &   12 &     --1.15 &   0.26 \\
    V246   &     --0.15 &   --2.82 &     6.80 &   --344 &   12 &     --1.35 &   0.29 \\

\enddata
\end{deluxetable}

\begin{deluxetable}{cc}                                                         
\tablewidth{0pt}                                                                
\tablecaption{Target catalog references\label{tab:targref}}                   
\tablehead{                                                                     
        \colhead{Code} & \colhead{Reference}                                    
}                                                                               
\startdata                                                                      
B   & Battistini et al.~(1980, 1987)\\                                          
BA  & Baade \& Arp (1964) \\                                                    
BoD & Battistini et al.~(1987, Table VI) \\                                     
DAO & Crampton et al.~(1985) \\                                                 
NB  & Battistini et al.~(1993) \\                                               
S   & Sargent et al.~(1977) \\                                                  
V   & Vete\u{s}nik (1962) \\                                             
\enddata                                                                        
\end{deluxetable}

\subsection{Numbers and Spatial Distribution}

\begin{figure}
\includegraphics[scale=0.6]{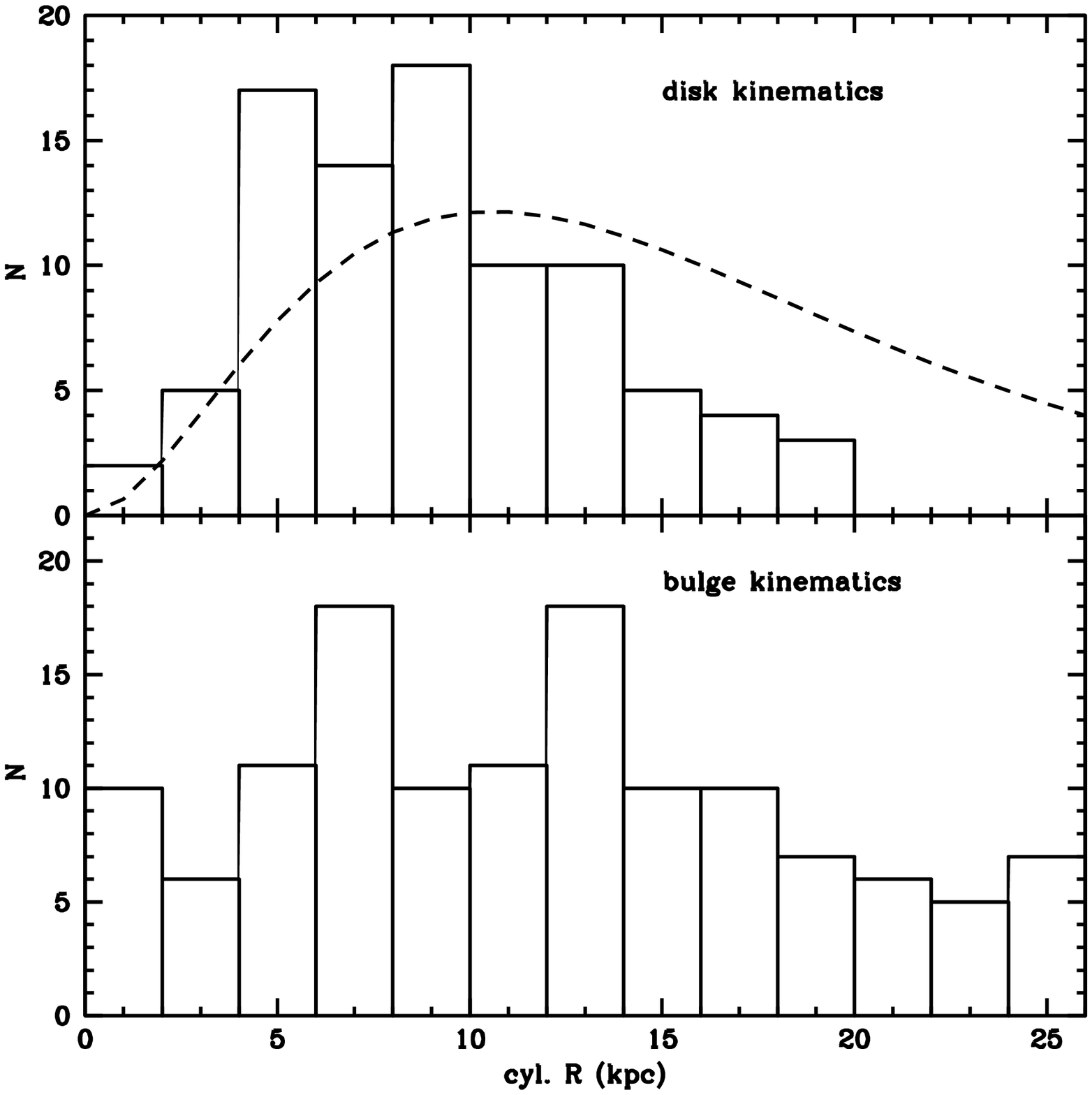}        

\caption{Histogram of deprojected cylindrical R values for both disk
(upper panel) and non-disk (lower panel) clusters (with absolute value
of residual less than and greater than 0.75 respectively). All
clusters found projected on the disk with velocity errors less than 20
km/s are plotted. The dotted
line in the top panel shows the fall-off of an exponential with scale
length 5.3 kpc (the distribution of the disk starlight), with
arbitrary normalization.
\label{cylRhist}} 
\end{figure}

First, we consider the distribution across M31's disk of the clusters with
disk kinematics. Figure \ref{xyplot} shows that the thin disk clusters are
found across the entire disk of M31. We can quantify this more exactly by
examining histograms of the distribution of deprojected cylindrical R for
clusters projected on the disk, shown in Figure \ref{cylRhist}. 
We find that 40\% of all clusters seen projected on the disk have disk
kinematics. 

We see from Figures \ref{yydist} and \ref{xxdist} that there are
significant numbers of known clusters \citep[from the compilation
of][]{barmby00} with small R whose velocities have not yet been
measured. Figure \ref{yydist} (lower panel) also shows that the
compilation of \citet{barmby00} is incomplete for small R. Clearly we
will need to revisit this question when the sample of globular
clusters with small R and velocity estimates has been better
explored. At this stage we note that it is possible that the
distribution of disk clusters is similar to the distribution of disk
starlight, although they appear somewhat more concentrated.  However,
we can see that there was a large disk in place at the epoch when
these globular clusters formed.

\subsection{Metallicity}

In the Milky Way, most globular clusters with disk kinematics are 
metal-rich, with [Fe/H]$>$--0.8 \citep{zinn85}, although
proper-motion measurements have shown that there are a few clusters
with [Fe/H]$<-0.8$ with disk-like orbits
\citep{cudhanson,dinescu99}. 
We know that kinematics are not as well-correlated with abundance
in M31 because \citet{perrett} showed that M31's
metal-rich clusters have only slightly more rotational support than the
metal-poor ones.  In fact we find that the metallicity distribution of the
clusters with disk kinematics and the rest are roughly similar: both groups
contain both metal-rich and metal-poor clusters.

\begin{figure}
\includegraphics[scale=0.7]{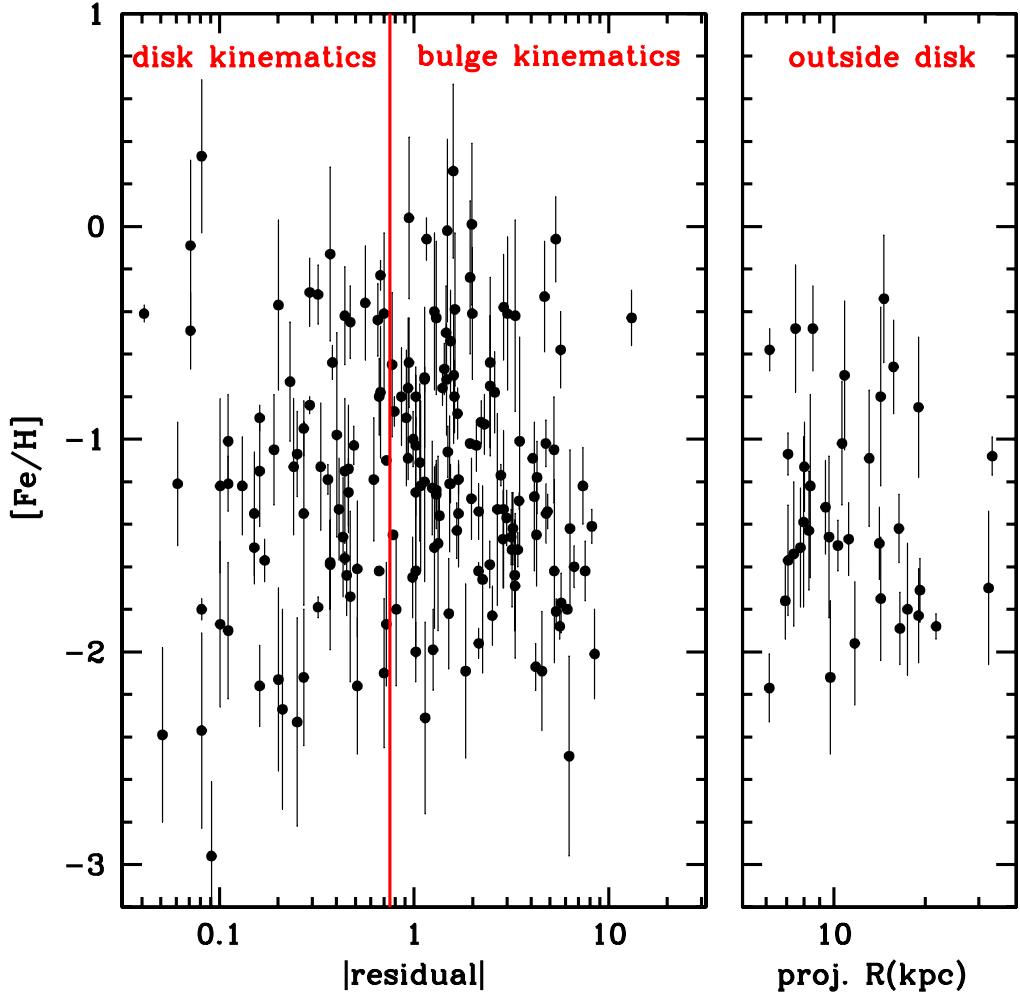}        

\caption{Metallicity distribution for, from the left, likely disk clusters
($|$resid$<0.75$), clusters projected on the M31 disk with bulge
kinematics, and clusters outside the boundary of the disk. Only
clusters with [Fe/H] errors less than 0.5 dex are shown.  Apart from
the larger spread in metallicity for the disk clusters (likely to have
been caused by larger errors in \fe\ measurement because of their
fainter magnitudes) there appears to be little difference between the
metallicity distributions of the two samples seen projected on the
disk. There is some evidence for an abundance gradient between the
clusters projected on the disk and those outside the disk.
\label{feresid}} 
\end{figure}

Figure \ref{feresid} compares the metallicity distribution of clusters that
are likely members of the thin disk subsystem with two other groups: clusters
projected on the disk whose kinematics are significantly hotter, and those
whose position is outside the disk altogether. The disk clusters (with
$|$residual$|<0.75$ have larger metallicity errors than the rest, probably
because of their fainter V magnitudes (see Fig. \ref{finalvmag}). However,
apart from the larger apparent spread in disk metallicities to both high and
low values, likely to be an artefact of the larger measurement errors, the
metallicity distributions of the two kinematical groups projected on the disk
look remarkably similar. 

In Figure \ref{feresid} we see that there are more objects with bulge
than with disk kinematics. Thus we need to consider the contamination
from the more numerous bulge clusters in the objects with small values
of the residual. Using the histograms of Figure \ref{fakeresidhist} as
a guide, we find that there are 28\% of bulge objects with
$|$residual$|<0.75$,  but only 8\% with $|$residual$|<0.25$. So, with
129 bulge and 88 disk clusters, we would expect 10 of the 38 clusters
with $|$residual$|<0.25$ to be interlopers from the bulge. We note
that the clusters with $|$residual$|<0.25$ still show the large range
in [Fe/H] that we remarked on for the larger group of disk clusters.

\begin{figure}
\includegraphics[scale=0.7]{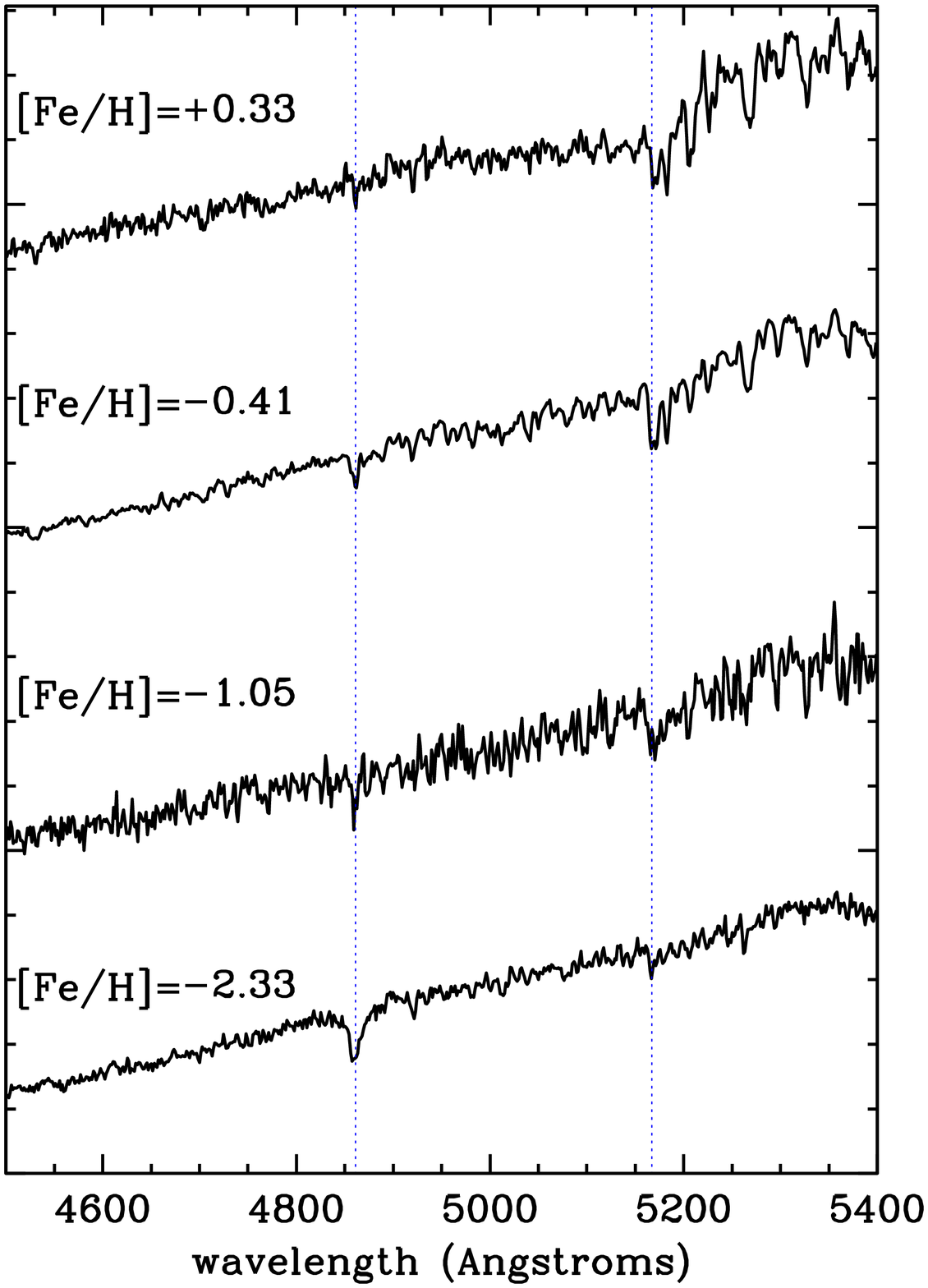}        

\caption{Spectra for likely disk clusters
 ($|$resid$|<0.75$). The clusters are, from the top, DAO70 (\fe=+0.33), B171
(\fe=--0.41), B335 (\fe=--1.05) and B327 (\fe=--2.33).
\label{kathyspectra}} 
\end{figure}

\begin{figure}
\includegraphics[scale=0.7]{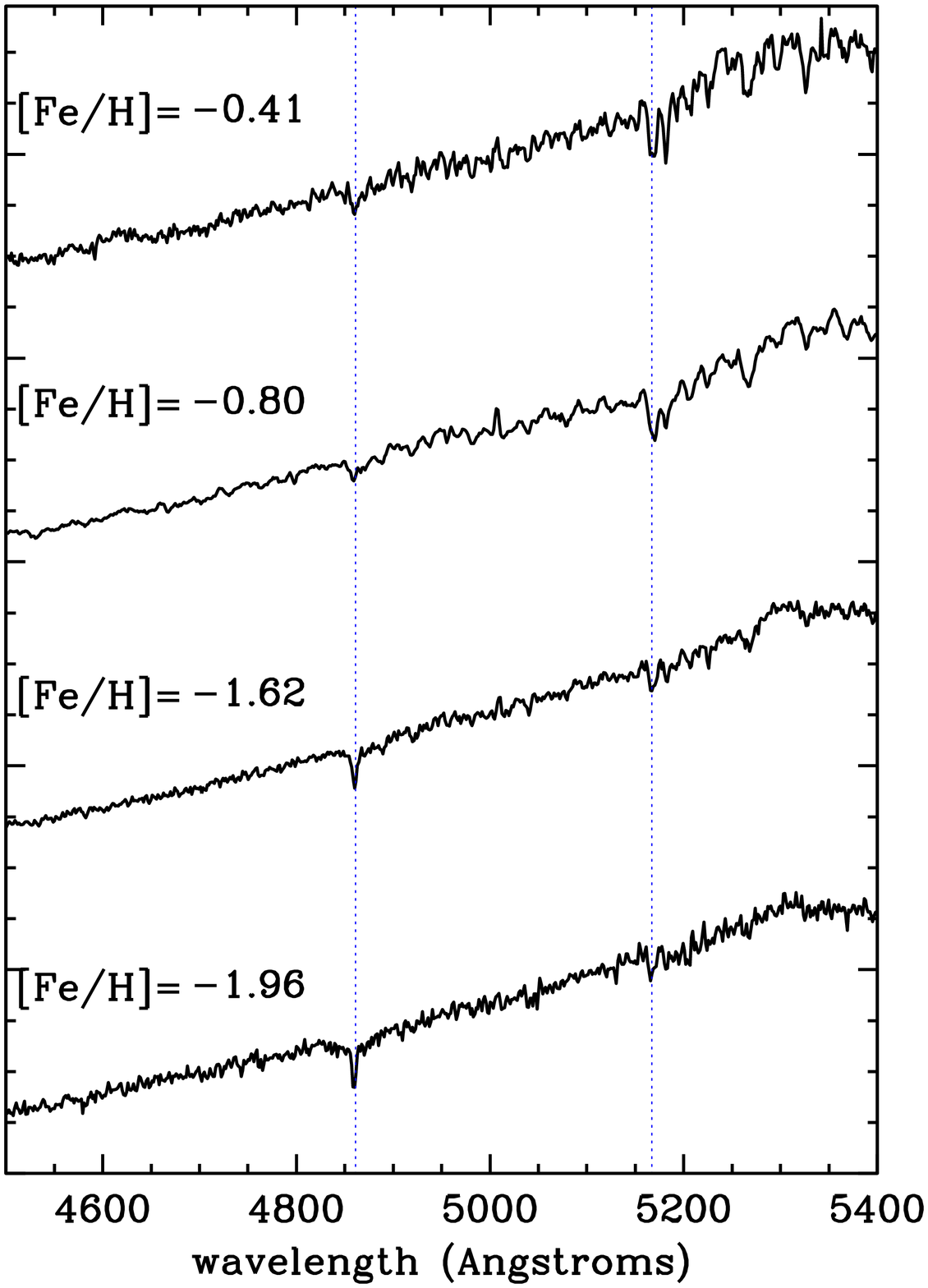}        

\caption{Spectra for clusters projected on the disk with non-disk kinematics
 ($|$resid$|>1$). The clusters are, from the top, B94 (\fe=--0.41), B127 
(\fe=--0.80), B135 (\fe=--1.62) and B311 (\fe=--1.96).
\label{kathyspectra2}} 
\end{figure}

Figures \ref{kathyspectra} and \ref{kathyspectra2} show the WYFFOS
spectra \citep{perrett} of four clusters from each of the disk and bulge groups. These
spectra confirm the results of Figure \ref{feresid}: both metal-rich
and metal-poor clusters can be seen in both groups.

The metallicity distribution of the thin disk globular clusters in M31
is consistent with an initial disk forming from very low-metallicity
gas and subsequent enriching up to solar abundance or even higher
during the globular cluster formation era. The metallicity
distribution of disk globulars in the Milky Way has led to suggestions
that the proto-disk gas was pre-enriched by star formation in the
spheroid \citep[eg][]{larson76}, but this assumption is not necessary
in the case of the M31 clusters.

While it is often risky to use metallicity as a rough measurement of
age, as star formation can proceed at different rates in a galaxy, we
can be fairly sure that the most metal-poor disk clusters with
\fe$<$--2.0 formed in the early stages of M31's disk formation. This
is because it would be very difficult for the gas that the disk
globular clusters formed from to remain un-enriched by the ongoing
star formation in M31's disk: both mass loss from high-mass stars and
supernova ejecta would mix into the existing disk gas, enriching it to
higher metallicity. In Figure \ref{xyplot} we show the spatial
distribution of both the disk globular clusters with \fe$>-2.0$ (red
filled circles) and of the most metal-poor disk globular clusters with
\fe$<-2.0$ (blue stars). The fact that the most metal-poor disk clusters
are found across much of M31's disk is another indication that M31 had
a reasonably large disk at early epochs.

\subsection{Cluster Reddening and Luminosity Function}

In the past, measurements of the luminosity function (LF) of the M31 
globular clusters have used only clusters outside the disk region of
M31 so that reddening from the disk does not bias the result. However,
the available data has increased significantly both in quantity and
quality in the past few years. \citet{barmby00} compiled photometry of
the M31 clusters in UBVRI and JHK, and used color-color relations from
the low-reddening Milky Way globular clusters to estimate the
reddening of over 300 M31 globulars. These cluster reddenings, kindly
made available by Pauline Barmby, allow us to compare the reddening
and absolute magnitude of the disk clusters to the rest.

\begin{figure}
\includegraphics[scale=0.7]{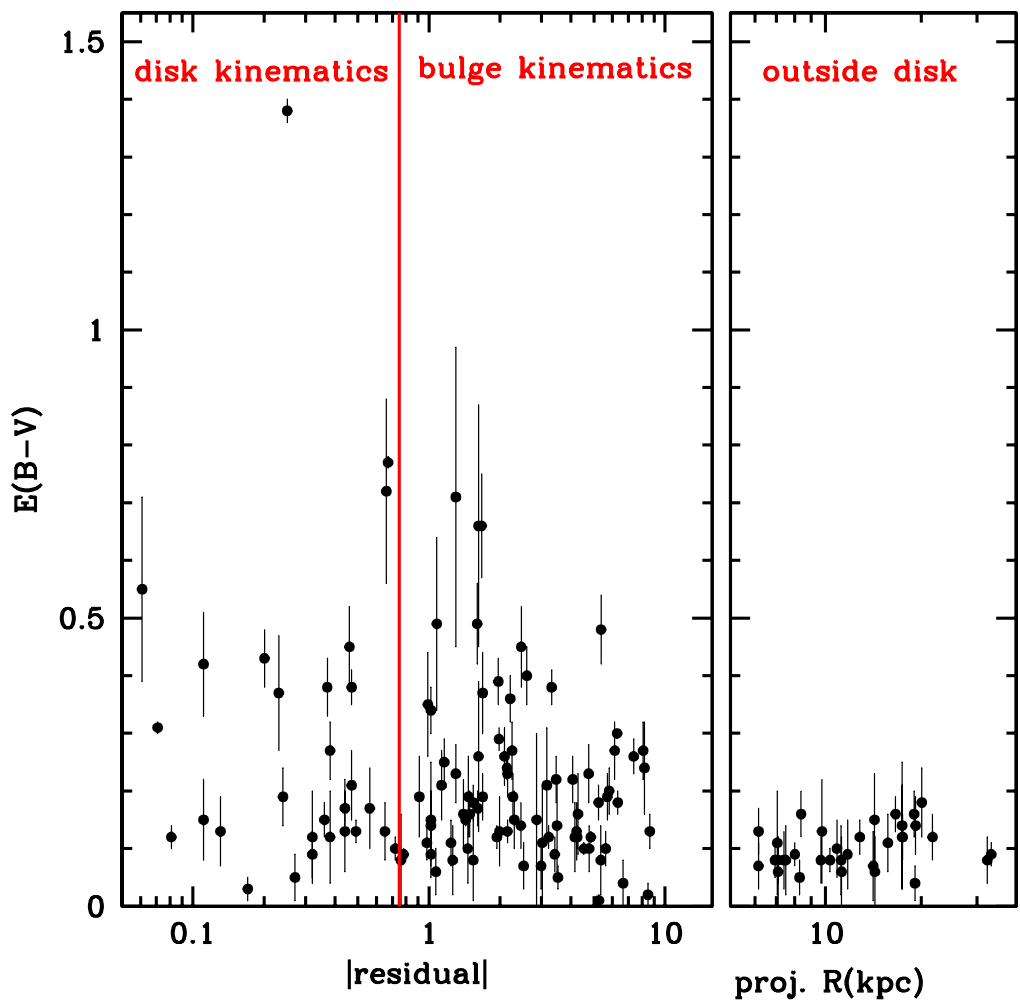}        

\caption{Variation of E(B--V) measurements from \citet{barmby00} for
clusters with reddening measurements classified as ``good'' by these
authors as a function of (left panel) the residual that indicates
likely disk membership and (right panel) projected distance from the
center for stars outside the disk. Within the errors, all clusters
outside the disk region have E(B--V) due to the Milky Way foreground
reddening only, while some clusters projected on the disk have much
larger values, as we would expect if they were within or on the far
side of the M31 disk dust layer.
\label{finalebv}} 
\end{figure}

We examine the reddening values for the clusters from the
\citet{perrett} sample with good velocities which also have reliable
reddening measures from \citet{barmby00}.  E(B--V) values are shown in
Figure \ref{finalebv}. First, it is comforting to note that the
clusters outside the disk region have values consistent with
foreground reddening from the Milky Way only.  Clusters projected on
M31's disk can be both in front and behind the disk reddening layer,
so we would expect the clusters in this region which are in front of
the M31 disk to have reddenings as low as the clusters outside the
disk, as is seen. However, other clusters (both with and without disk
kinematics) have much higher reddening values, suggesting that they
are either in or behind the M31 dust layer. 

However, before we proceed further, it is necessary to calculate the
number of interlopers from the bulge that are likely to be in the disk
region of the Figure. There are 98 clusters with $|$residual$|>0.75$
and 46 with $|$residual$|<0.75$ in Figure \ref{finalebv}. With almost
twice as many bulge as disk clusters, the interloper problem becomes
severe. For the bulge model of Table \ref{modelparams}, we would
expect about 27 bulge interlopers with $|$residual$|<0.75$ and 8 with
$|$residual$|<0.25$, compared with 46 and 16 clusters respectively
observed. Thus approximately 50\% of the objects with disk kinematics
in Figure \ref{finalebv} are likely to be interlopers from the
bulge. Worse, this number may vary significantly depending on the exact
kinematics of the bulge clusters and on small number statistics. So,
although we will proceed to calculate the luminosity function for the
bulge and disk clusters, these results should be viewed as preliminary
only. We need a larger sample of disk clusters with accurate values of
E(B--V) before we can calculate the luminosity function of the disk
clusters with any reliability.

\begin{figure}
\includegraphics[scale=0.7]{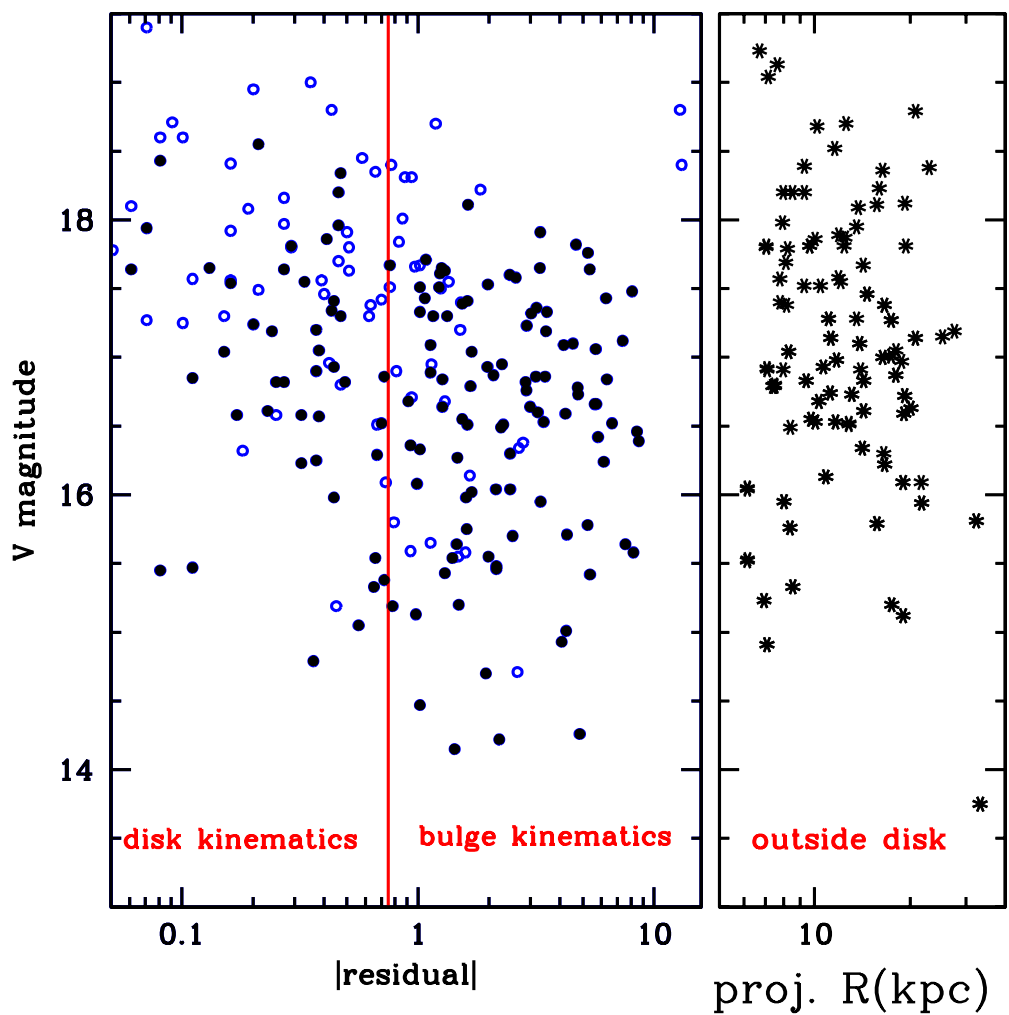}        

\caption{Variation of V magnitude (not corrected for reddening) for
  clusters as a function of (left panel) the residual that indicates
  likely disk membership when clusters are projected on the disk, or
  (right panel) projected distance from the center for clusters
  outside the disk. In the left panel, closed symbols are clusters
  from \citet{perrett} which have reddening estimates from
  \citet{barmby00}, while open symbols have good velocity estimates
  from \citet{perrett} but no E(B--V) estimate. In the right panel we
  have used all clusters from the compilation of \citet{barmby00}
  whose positions are outside the disk region.  It can be seen that
  approximately half the clusters with disk kinematics have no E(B--V)
  estimate.
\label{finalvmag}} 
\end{figure}

Since discovery of disk clusters is likely to be more difficult
against the bright and variable background of the disk, we need to see
whether the samples have similar limiting magnitudes before we compare
their luminosity functions. Figure \ref{finalvmag} shows the
distribution of V magnitude (uncorrected for reddening) for the
cluster sample as a function of disk membership residual or projected
distance from the center.  We note first that the clusters with
multicolor photometry available (needed to provide a reddening
estimate, shown with solid circles in Figure \ref{finalvmag}) are less
likely to belong to the disk. Clearly there is a need to obtain good
photometry in a number of passbands for more of the clusters projected
on its disk.  Clusters outside the disk are not well represented in
the velocity sample of \citet{perrett} because they were not able to
obtain spectra there due to poor weather. Thus we have simply used the
clusters from the compilation of \citet{barmby00} to show the V
magnitude measures of clusters outside the disk region.

It can be seen in Figure \ref{finalvmag} that the clusters outside the
disk will have a luminosity function that extends fainter than the LF
of the clusters inside the disk region for two reasons. We can safely
assume that all the clusters outside the disk region have the same
reddening (see Figure \ref{finalebv}) and so can use the entire sample
in the right panel of Figure \ref{finalvmag} for constructing a
LF. However, inside the disk region, we will only be able to
use clusters with actual reddening estimates (shown with filled
symbols) in making the LF for the disk region. These clusters have
brighter V magnitudes than the clusters outside the disk, and because
they have higher reddening on average they will have even brighter
absolute magnitudes. However, we can compare the
luminosity functions of clusters {\bf projected on the disk} with disk and
non-disk kinematics, because the limiting magnitudes of these two samples
are similar. 

\begin{figure}
\includegraphics[scale=0.6]{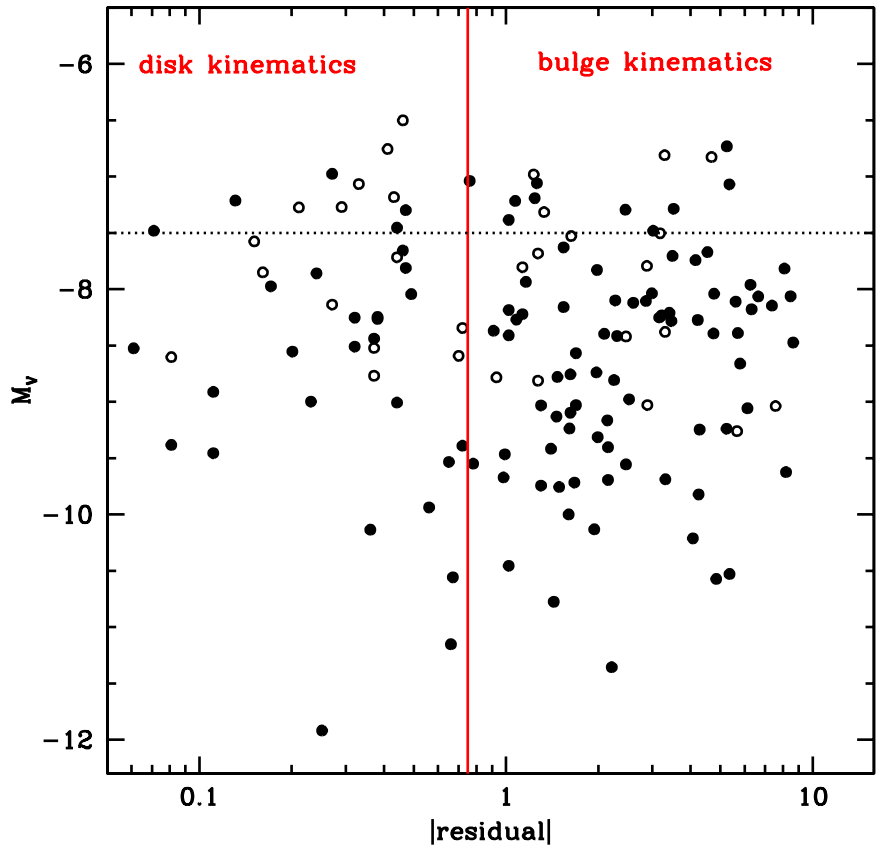}        

\caption{Absolute magnitude $M_V$ as a function of the disk membership
  residual for all clusters from \citet{perrett} with reddening estimates from
  \citet{barmby00}. Closed symbols denote clusters whose reddening
  estimate was classified as ``good'', open symbols all other clusters
  with a reddening estimate from \citet{barmby00}. 
\label{finallf}} 
\end{figure}

Figure \ref{finallf} shows that, with our current smallish sample of
20--40 disk clusters with \mbox{E(B--V)} estimates, there is no perceptible
difference between the luminosity functions of clusters with and
without disk kinematics.  We are working to obtain E(B--V) estimates
for a larger sample of clusters projected on the disk. At present
we find that there is no  strong evidence for a
difference in luminosity function with kinematics.

If this result is borne out with a larger sample of disk clusters,
thus avoiding much of the problems with bulge contamination in the
disk area, we
could conclude from the similarity of luminositity functions
and metallicity distributions that there are unlikely to be large
differences in mass or age between the two samples ..... unless the disk
clusters are both less massive and younger and the two effects cancel.
We will revisit this question in a future paper when we will use
integrated colors and spectra to measure ages more accurately.

Even at this preliminary stage, however, it is clear that the
luminosity function of the thin disk globulars in M31 is quite
different from that of the thin disk old {\bf open} clusters in the
Milky Way. The open clusters with ages greater than 1 Gyr have a mean
$M_V$ of --3.5 \citep{battinelli}, in contrast to the Milky Way or M31
globular clusters whose mean $M_V$ is close to --7.5.

\subsection{Disk heating mechanisms and how they affect globular clusters}

Young disk stars in the Milky Way have a low velocity dispersion (of
order 10 km/s). Their orbits acquire additional energy via encounters
with inhomogeneities in the disk potential such as spiral arms and
giant molecular clouds \citep{jenkins}, and old disk stars are found
to have velocity dispersions of order 30--40 km/s. The mass of an
individual globular cluster is close to that of a molecular cloud but
significantly smaller than a spiral arm. It is possible that the
heating processes will be different in interactions between objects of
similar mass.

It is thought that interactions between stars and spiral arms heat mostly
the radial and azimuthal velocity components, while molecular clouds
affect the vertical velocity \citep{jenkins}. So we might expect the
vertical velocity dispersion of the disk globular clusters to be lower
than that of the disk stars if cluster-molecular cloud interactions
are less effective, but the R and $\Phi$ dispersions of clusters and
stars to be similar. Because M31 is close to edge-on, the line of
sight velocity is dominated by the R and $\Phi$ components, so it is
not surprising that the disk globular cluster velocity dispersion is
close to that predicted by the Bottema relations for disk stars.

Several groups are currently studying the kinematics of PNe in M31
\citep[eg][]{denise,halliday} and this will allow a direct comparison to be made
between the velocities of old disk stars and of globular clusters in M31. If
this result holds up, it will provide a confirmation of these theories of
secular disk heating.

\subsection{Mergers in M31}

A more substantial form of disk heating (or destruction) occurs during
a minor or major merger. When an object of order 10\% of the mass of a
disk galaxy is accreted, we term it a minor merger, while ``major
merger'' refers to the situation when the two colliding galaxies are
of roughly equal mass.  A minor merger transfers some of the orbital
energy of the satellite to the disk stars, and the disk becomes both
thicker and dynamically hotter, but still remains intact
\citep{quinn86,walker}. Such a merger was thought to have occurred in the
Milky Way about 10 Gyr ago \citep{kenaraa,edvardsson}.  A major merger
will destroy an existing disk completely, with an elliptical galaxy
being the most likely final outcome.

When thick disks were first discovered in external galaxies
\citep{vdks1,vdks2} it was noted that they were found only in galaxies
with substantial bulges.  The reason for this correlation is not
clear, although it is possible that galaxies with substantial bulges
experience more minor mergers than late-type systems
\citep[eg][]{silkwyse}.  Using this correlation, we would have
expected to see a thick disk in M31. However, we have found no strong
evidence of this either in PN kinematics \citep{denise} or in globular
cluster kinematics\footnote{The claim by \citet{ata} that M31 has a
thick disk is based only on a 
color-magnitude diagram of a field several kpc from M31's major
axis. Given the existence of a warp in M31's disk \citep{wk88} and the
lack of kinematical evidence for a thick disk, we think that it is
more likely that their relatively metal-rich component is related to
the warped thin disk}. In fact, we can make an even stronger
statement: if M31 had experienced a merger, whether major or minor,
since its disk globular clusters were formed, they too would have been
heated by the merger and would have a significantly larger velocity
dispersion than observed. It has been merger-free since their
formation.

\citet{tombrown} found, using very deep HST ACS photometry of a field
11 kpc out on the SW minor axis, that there is a population of younger
stars in this field. Their initial estimate for the age of the younger
group was 6--8 Gyr, but this may not be unique due to the complexities
of fitting CMDs for mixtures of populations. They suggest several
scenarios to explain the relatively high metallicity and large age range of
stars in this field. One postulates an interaction between the M31 disk and a
roughly equal-mass companion galaxy 6--8 Gyr ago, which either placed
disk or satellite stars 11 kpc above the plane. As we have discussed,
the coldness of the thin disk clusters rules out any such merger since
the clusters were formed.

The M31 globular clusters do not in general have accurate age
estimates: until the ACS observations of \citet{tombrown}, {\bf no}
cluster had a color-magnitude diagram which reached to the main
sequence turnoff, and in fact the disk globular clusters as a class
are even less well-studied, with no HST observations which reliably
detect the horizontal branch. Thus we are forced to depend on
integrated photometry or spectroscopy to derive age estimates.

Early work by \citet{burstein}, using the Lick spectroscopic indices,
claimed that the M31 clusters were younger than the Milky Way ones
because of their stronger H$\beta$ indices. However, later workers
were in general unable to reproduce their results
\citep{bhuchra1,bhuchra2,hbrodie}, and most recent estimates, using
high-quality optical or UV spectra, conclude that there is currently
little evidence for a difference in mean ages between the two globular
cluster populations \citep{bohlin,ruth} \footnote{There are preliminary
indications that the metal-rich clusters are brighter on average than
the metal-poor ones \citep{barmbylf}, and one possibility is that the
metal-rich clusters are younger, but this needs to be confirmed with a
larger and more complete sample.}. The disk globular clusters
pose the additional problem that a number of them are significantly
reddened, so both E(B--V) and age will need to be derived reliably.

Thus, although it seems unlikely that all the M31 disk globular clusters are
younger than 6--8 Gyr, we cannot rule it out at this time. Thus there are two
possibilities: 
\begin{itemize}
\item[(1)] All the disk globular clusters in M31 are younger than 6--8
Gyr. In this case there is no conflict between our results and the
equal-mass merger scenario of \citet{tombrown}. However, it seems
unlikely that it was possible to form clusters with metallicities as
low as \fe = --2.0 {\bf in M31's disk} as recently as this, as any star
formation in the galaxy would result in pre-enrichment of the gas from
which the cluster formed. The survival of a significant mass of gas
($10^6$ solar masses or more) at such low abundance in the disk for a
significant  time after star formation started seems very
unlikely.
\item[(2)] The disk globular clusters are of similar ages to the Milky
Way globulars. In this case, their kinematics rule out the equal-mass
merger suggested by \citet{tombrown}. Any satellite interactions
responsible for placing younger stars 11 kpc from M31's disk must have
been with satellites of sufficiently low mass to leave the disk
undisturbed. We know that a satellite of mass 10\% of the disk mass
will heat it significantly, so any interaction must have been with a
satellite whose mass was a few percent or smaller, such as M32 whose
current mass is 2\% of M31's disk mass. 

Using the relation between metallicity and mass for Local Group
galaxies \citep{marioaraa}, we find that most of the dwarf galaxies in
the Local Group have \fe $\lesssim -1.0$, excepting more massive satellites
such as the LMC, which would significantly damage a galaxy disk when
accreted. The mean metallicity of stars in the ACS field ([Fe/H]$\sim
-0.6$) is interestingly high in this context.  If a smaller satellite
with mass similar to M32 or NGC 205 were accreted, we would expect to see a
combination of ages and metallicities not explored by
\citet{tombrown}: an old population with metallicity extending across
the entire range of the M31 globular clusters, plus a younger {\bf
relatively metal-poor} population accreted from the infall of a
smaller satellite.
\end{itemize}

There have been interesting recent detections of stellar streams in the
outer regions of M31 \citep{ibata,annette}. While these features are
likely to be tidal in origin and produced by satellite accretion, we
emphasise that any satellite progenitor is likely to be {\bf several
orders of magnitude} less massive than needed to substantially heat
the M31 disk. For example, \citet{ibata} quote an average surface
brightness of $\mu_V$ = 30 mag/arcsec$^2$ for the giant stream near
the southeastern minor axis, and an absolute magnitude \mbox{XS$M_V$ =
--14}. For comparison, the Fornax dwarf spheroidal galaxy has $M_V$ =
--13.2 and a mass of order 10$^7$ solar masses \citep{marioaraa}, more
than two orders of magnitude too low in mass to damage a large disk
such as M31's, while M32 has an absolute magnitude of $M_V$=--16.7 and
a mass of approximately 2\% of the M31 disk mass \citep{marioaraa},
still too low in mass to do significant damage.

\section{Discussion: Disk Formation and Evolution}
\label{s:discussion} 

\subsection{Comparison with Other Galaxies: the Local Group}

As discussed above, the only clusters in the Milky Way with thin disk
kinematics are the open clusters, which are several orders of
magnitude less luminous than the M31 globular clusters. 

Is it possible that a population of thin disk globular clusters exists
in the Milky Way but has remained undetected because of extinction by
dust? The 2MASS survey has the capacity to detect Galactic objects
which are highly reddened, although it is limited by confusion for
objects close to the plane.  It has only produced 2 candidate globular
clusters, at (l,b) of (10,0) \citep{hurt2mass}.  The SIRTF satellite,
and in particular its planned survey of the galactic plane region,
will be able to make a much more sensitive search for these objects.
However, while it is possible that there still exists a population of
thin disk globular clusters close to the galactic center, it is much
less likely that such a population would remain undetected close to
the Sun. We have already noted that the M31 disk globulars are found
across its entire disk. It is thus unlikely that the Milky Way has a
similar population.

However, all but a few of the Milky Way globular clusters are older
than 10 Gyr, when we believe that the Milky Way suffered a minor
merger which heated the stars then in its thin disk into a thick
disk. Since any globular clusters in the then thin disk would have
felt the same gravitational forces and would also have become part of
the thick disk, we would not expect to find any globular clusters
older than 10 Gyr in a thin disk configuration in the Milky Way.

It is interesting to compare the spatial distribution of the most
metal-poor globular clusters in the Milky Way's thick disk with that
of the metal-poor globulars in M31's thin disk, since presumably both
systems started as a thin disk. There are four Milky Way globulars
with thick-disk orbits: NGC 6121, NGC 6254, NGC 6626 and NGC 6752
\citep{cudhanson,dinescu99}. They have galactocentric radii of 5.9,
4.6, 2.7 and 5.2 kpc respectively \citep{harris96}, which at first sight
looks significantly smaller than the mean radii of the disk globular
clusters in M31 (8--10 kpc). However, when we recall that the Milky
Way disk is not only less luminous but also smaller than M31's, with a
scale length of 2--3 kpc \citep{kent91,drimmel}, compared to 5.3 kpc
for M31 \citep{wk88}, then we find that the early disk in both
galaxies was probably of a similar size compared to their final scale
lengths.

There are two other disk galaxies in the Local Group: the LMC and
M33. Both are significantly less massive than the Milky Way and M31.
While there are a small number of true globular clusters in the LMC
\citep{nickaraa} which all have disk kinematics, it is not clear at
this stage that there are any similarly old clusters with disk
kinematics in M33. \citet{schommer,chandar} show that M33 clusters with
ages greater than $\sim$1 Gyr have halo kinematics. It is possible
that a study of the oldest M33 clusters with similarly high velocity accuracy
to the \citet{perrett} study would show more kinematic substructure in
this population.

\subsection{Galaxies Outside the Local Group}

If a population of globular clusters with disk kinematics has existed
in the nearest large spiral and remained undetected until now, it
is quite possible that there are similar populations in other
nearby spirals. 

It is interesting to note in this context the WFPC2 survey for globular
clusters in the disk regions of two nearby edge-on spirals by
\citet{zepf5907}. The two spirals, NGC 4565 and NGC 5907, have similar disk
luminosities and rotation curve amplitudes, but quite dissimilar bulge-to-disk
ratios and stellar populations. NGC 4565 is often used as an edge-on analog of
the Milky Way, and has a large bulge (which has been found to be a strong bar
viewed close to end-on, Kuijken et al, in preparation) and a bright thick
disk with scale height 2 kpc \citep{hlm1999}. NGC 5907, by contrast, has a
small, almost unresolved bulge, no bright thick disk such as that seen in the
Milky Way or NGC 4565, and a faint flattened halo \citep{mbh5907,sackett5907}.

Basing our expectations on the Milky Way, with globular clusters
associated with the thick disk (or perhaps the bulge) and the halo, we
would expect a survey for globular clusters in the disk regions of NGC
4565 and 5907 to produce very different numbers of clusters, because
of the much stronger thick disk in NGC 4565. Surprisingly, this is not
the case. \citet{zepf5907} find that, within the errors, NGC 5907 and
NGC 4565 have the {\bf same} number of globular clusters in the
disk-dominated regions they studied.  This suggests that there may be
a population of globulars associated with the thin disk in these
galaxies, as it is the only stellar population with similar
properties in this pair. Velocity measurements for these
globular clusters will be an useful check of this suggestion.

\subsection{The Epoch of Disk Formation}

We have shown that M31 has a significant population of globular
clusters associated with its {\bf thin} disk. It is clear that these
are at least moderately old (of order 10 Gyr).  In addition, the very
cold kinematics, which would have been destroyed by even a minor
merger, show that the disk has been in place in M31, undisturbed,
since the clusters were formed.

This observation provides an interesting complement to the detection
of rotating systems at redshifts larger than 1
\citep[eg][]{vandokkum,erb}. Their spatially resolved disk kinematics
for high-redshift galaxies show that there were some reasonably large
disks in place at that time, and the detection of large disk-like
galaxies with symmetrical, exponential surface brighness distributions
in the rest-frame optical at z $\sim$2 by \citet{labbe} also hint at
an early epoch of disk {\bf formation}.  But we do not know what the
ultimate fate of these disks will be. In fact, \citet{labbe} note that
their disk-like galaxies are strongly clustered (no similar galaxies
are found in the HDF-N) and suggest that they may be the progenitors
of cluster S0 galaxies. It is possible that disk formation is a common
process, but that many of the disks that form early do not survive.

The areas probed by the very deep surveys needed to identify disks at
high redshift are still small, although this is a very active area
where several groups are making progress. A combination of low-redshift
kinematical studies of nearby disk globular cluster systems such as
this paper, and a more complete census of disk galaxies at higher
redshift, will allow us to understand the era when the thin disks
that we observe today began to form.

\section{Summary}

We have discovered a new population of globular clusters in M31, with
thin disk kinematics. These clusters are spread over much of M31's
disk and have a metallicity distribution similar to the entire M31
globular cluster population, ranging from \fe\ below --2.0 to above
solar. While we do not have accurate measures of the age of M31
globular clusters yet, there is little indication at this point that they
are significantly younger than the Milky Way globulars. More accurate
age measures will constrain the formation epoch of the M31 disk, and
several groups are working actively in this area \citep[][Beasley,
Brodie, Forbes, Huchra \& Barmby, in preparation; Puzia, Perrett, \&
Bridges, in preparation]{ruth}. M31 is likely to have had a large disk
in place at early times.

The existence of such a dynamically cold system of presumably very old
objects places interesting limits on the accretion of satellite
galaxies since the clusters were formed: no minor merger with a
satellite of mass more than $\sim$10\% of M31's disk can have occured
since that time, because it would have heated the clusters into a
dynamically hotter system. This makes the suggestion of
\citet{tombrown} that M31 underwent an equal-mass merger 6--8 Gyr ago
less likely, and points to the urgent need for better age estimates
for the disk globular clusters. Accretions such as the one which
formed the giant tidal stream discovered by \citet{ibata} would have
involved a much lower mass satellite which would leave the M31 thin
disk unscathed.

Although the Milky Way is unlikely to have a similar population of
thin disk globulars, there are intriguing suggestions that there may
be similar groups of clusters in some nearby galaxies such as NGC 5907
and NGC 4565, which need to be followed up with velocity
data. Velocity studies of some face-on disk galaxies would also be
interesting, as we would expect the vertical velocity dispersion of a
population of thin disk globulars to be very small because of less
efficient secular heating by giant molecular clouds.

A search for such populations in nearby galaxies, and better data on
the ages of the M31 globular clusters, will provide an important
counterpoint to observations of large disk galaxies at redshifts of
order 2: while we are beginning to detect the signatures of such disks
at high redshift, we are not able at this point to determine whether
they would be disrupted as the galaxies evolve.


%% Included in this acknowledgments section are examples of the
%% AASTeX hypertext markup commands. Use \url without the optional [HREF]
%% argument when you want to print the url directly in the text. Otherwise,
%% use either \url or \anchor, with the HREF as the first argument and the
%% text to be printed in the second.

\acknowledgments

We would like to thank Ken Freeman for his helpful explanations of
rotation curves and disk heating, and Taft Armandroff, Gary Da Costa,
Ortwin Gerhard, Michiel Kreger, Chris Mihos and John Norris for other
useful discussions and suggestions. Much of this work was done while
HLM and PH were visitors at the Kapteyn Institute, Groningen, and
Sterrewacht Leiden: it is a pleasure to thank both institutes for
their hospitality. HLM and DHK acknowledge the support of NSF CAREER grant
AST-9624542 and AAP Fellowship AST-0104455.



\appendix

\section{Asymmetry in kinematics: reddening and completeness of cluster sample}

The cold kinematics of the disk globular clusters suggest that
their scale height is small. If so, we might expect a number of these
clusters to be within the dust layer of M31's disk.  If these clusters
are detected at all, they will have larger reddening values than
clusters in front of the disk. However, another possibility is that in
regions where the disk reddening is particularly high, clusters in and
behind the dust layer may not have been detected at all. If this is
the case, the kinematical signature of the disk will be less clear in
these regions because significant numbers of disk clusters will be
obscured.

The recent study of M31 cluster detection and completeness of
\citet{bh01} using the HST archive gives us the ability to test this
hypothesis. Barmby and Huchra searched the HST archive for WFPC2
images of M31, finding a total of 157 fields, many of which are in
disk regions.  As well as recovering 82 known globular clusters, they
found 32 new globular cluster candidates, and used the
location of the newly discovered clusters to quantify the
incompleteness of the sample.  Perhaps surprisingly given M31's
closeness, \citet{bh01} show that their sample begins to show
incompleteness at V=17, which corresponds roughly to the peak of the
GCLF at the distance of M31 \citep{harris96}. Thus a reddening of a
few tenths of a dex could well lead to a significant change in the
numbers of clusters known in that region.

\begin{figure}
\includegraphics[scale=0.6]{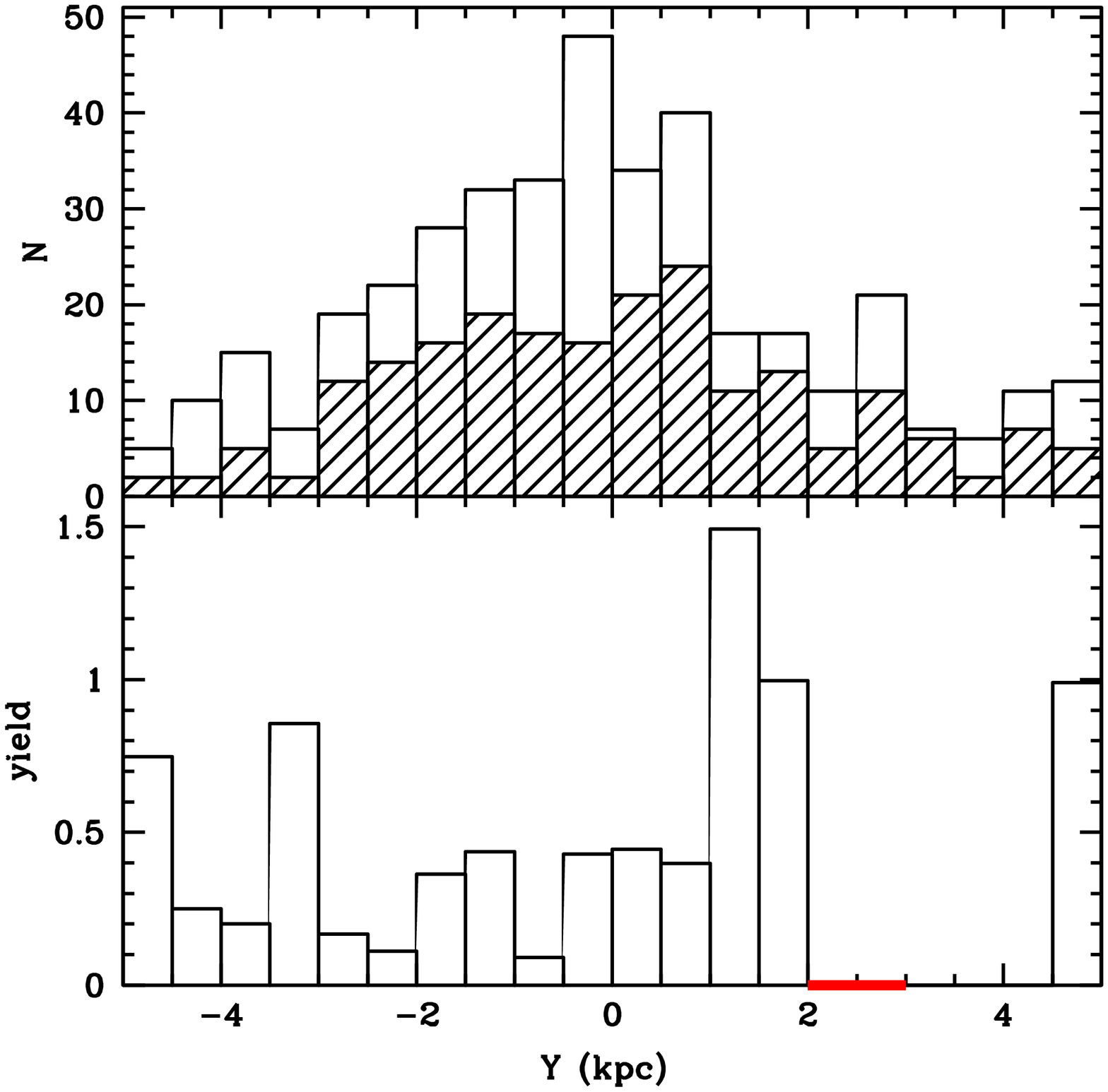}        

\caption{The top panel shows the distribution in Y (distance from the
major axis) of all globular clusters from the \citet{barmby00}
compilation (unshaded histogram) with the subset of this sample which
have velocity errors less than 20 km/s in \citet{perrett} shown
shaded. There are fewer clusters known with positive Y than negative
Y.  The bottom panel shows the ``yield'' of new clusters found by
\citet{bh01} in their study using the HST archive. We define the yield
as the number of new clusters discovered per WFPC2 field studied in
this range of Y. Regions with a thick solid line along the X axis of
the bottom histogram had no WFPC2 fields taken, so we have no
information about the completeness of the sample there.
\label{yydist}} 

\end{figure}

\begin{figure}
\includegraphics[scale=0.6]{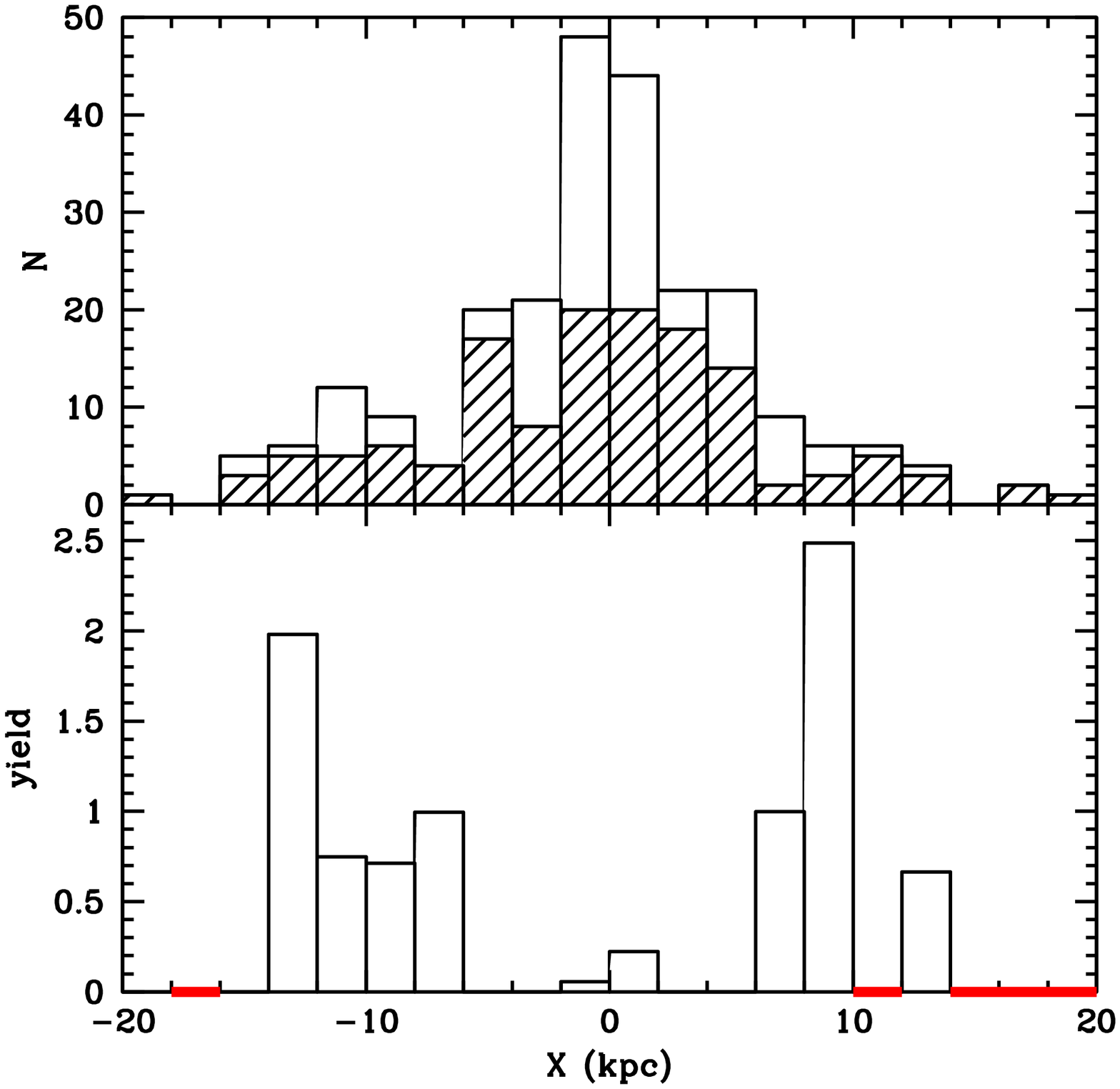}        

\caption{The top panel shows the distribution in X (distance along the
major axis) for $|Y|<$2 kpc of all globular clusters from the
\citet{barmby00} compilation (unshaded histogram) with the subset of
this sample which have velocity errors less than 20 km/s in
\citet{perrett} shown shaded. There are slightly fewer clusters known
with positive X than negative X. The bottom panel shows the ``yield''
of new clusters found by \citet{bh01} in their study using the HST
archive, defined in the same way as in the caption of Figure
\ref{yydist}.
\label{xxdist}} 
\end{figure}

In the top panel of Figure \ref{yydist} we show histograms of the Y
distribution of (a) the clusters in \citet{barmby00} and (b) the
subset of these clusters with velocity errors less than 20 km/s from
\citet{perrett}, which are the clusters our kinematical analysis is
based on.  It can be seen that there are fewer clusters known with
positive than with negative Y in both the \citet{barmby00} and the
\citet{perrett} samples. Figure \ref{xxdist} shows the X distribution
of clusters for values of $|Y|<$ 2 kpc (where we see the disk
signature most clearly) -- it can be seen that there are also fewer
clusters known with positive than negative X.

We note at this point that the quadrant with the fewest clusters (X
and Y both positive) is also the one with the lowest Galactic
latitude. This suggests that foreground extinction from the Milky Way
may be contributing to the incompleteness of the M31 cluster
sample. However, the small spatial range over which we see pronounced
asymmetries in the Y distribution corresponds to less than a degree of
galactic latitude. M31 is located at Galactic latitude b=--21.6. 2 kpc
at the distance of M31 corresponds to an increase of only 0.1 degrees
in b on the minor axis. Because of this it is unlikely that
foreground reddening is the major cause of this effect.

Using the new HST cluster candidates, we have quantified the
completeness of the cluster sample as follows. We first count the
number of WFPC2 fields in a given range of X or Y (counting fractional
fields in the few cases where 2 or more WFPC2 fields overlap). We then
count the number of new clusters detected in the same range of X or Y
(including candidate classes A, B and C from \citet{bh01}) and divide
the number of new clusters by the number of WFPC2 fields to give the
yield of new clusters per WFPC2 field. The bottom panels of Figures
\ref{yydist} and \ref{xxdist} show this yield for the X and Y
distribution of clusters. It can be seen that the regions where
clusters appear missing in the top panel are the same regions where
the incompleteness of the sample is greatest.

\begin{figure}
\includegraphics[scale=0.6]{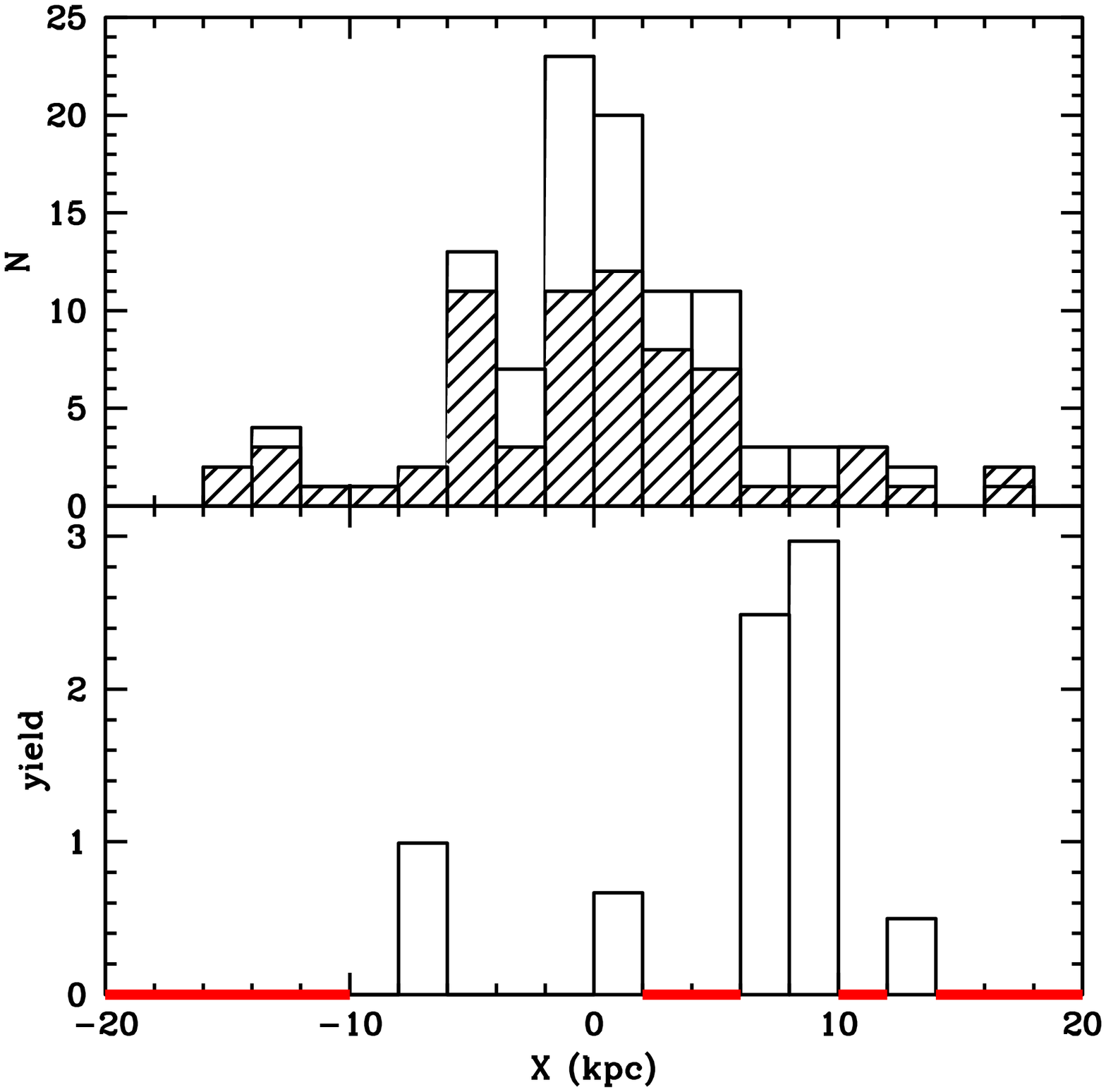}        

\caption{The top panel shows the distribution in X for clusters with
  positive Y of all globular clusters from the \citet{barmby00} compilation
  (unshaded histogram) with the subset of this sample which have
  velocity errors less than 20 km/s in \citet{perrett} shown shaded.
  The bottom panel shows the ``yield'' of new clusters found by
  \citet{bh01} in their study using the HST archive, defined in the
  same way is in the caption of Figure \ref{yydist}.  The quadrant
  with both positive X and Y has the largest incompleteness in our
  sample.
\label{xxdistypos}} 
\end{figure}

A particularly striking result is found for the side of the galaxy
with the fewest known clusters -- the X distribution for positive Y
only. This is shown in Figure \ref{xxdistypos}. For positive X, in all
regions where WFPC2 data were available, the sample is incomplete,
with yields of up to 3 new clusters per WFPC2 field.

We conclude that the region with  positive X and Y  has a less
complete cluster sample than the other quadrants of the galaxy.
We now examine the cluster kinematics for $|Y|<2$ kpc, split into
regions of positive and negative Y. If the cluster incompleteness
found above has led to fewer disk objects being found in this region
because they are hidden within the disk reddening layer, this should
be reflected in the kinematics.

\begin{figure}
\includegraphics[scale=0.7]{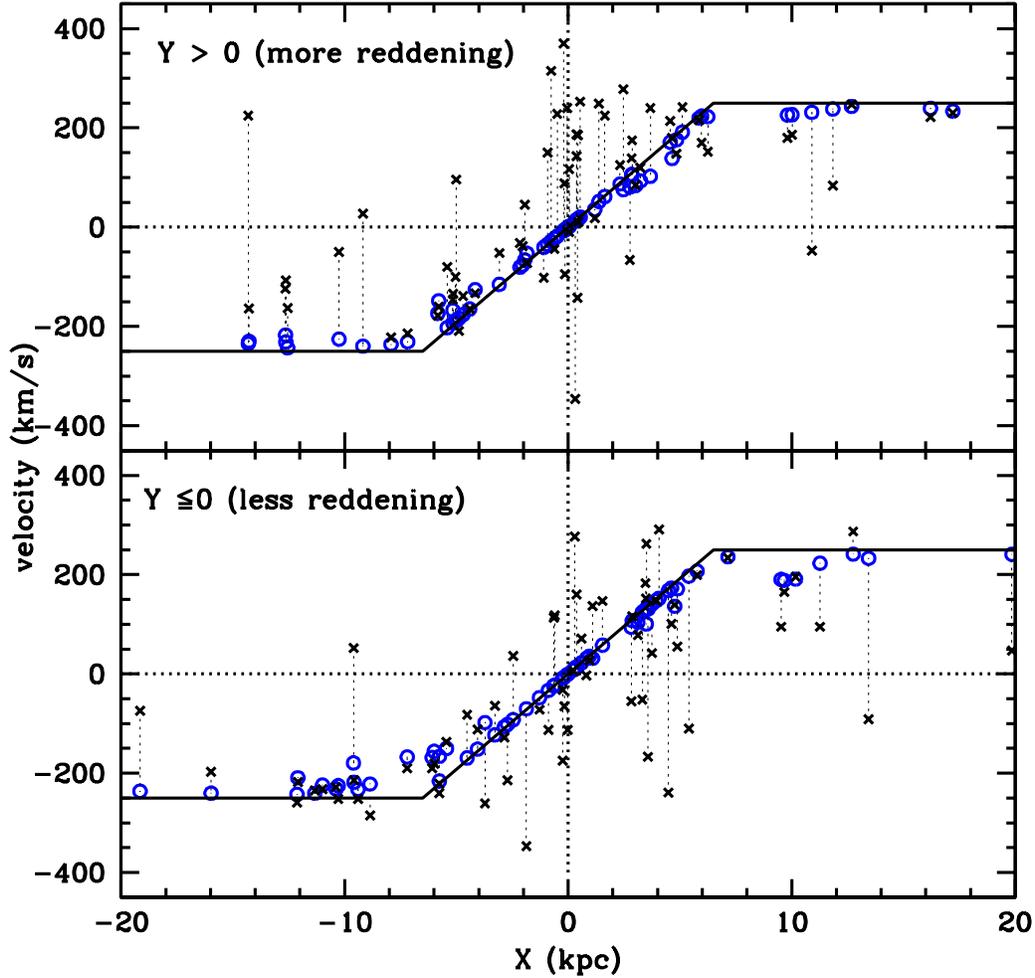}        

\caption{Predictions of a completely cold disk model (open symbols)
for clusters with velocity errors less than 20 km/s and $|Y|<2$
compared to actual velocities (filled symbols). A dotted line connects
the actual and the model velocity for each position.  The \citet{kent}
rotation curve is shown for comparison in each panel.  In the upper
panel we show clusters from the side of the galaxy where samples are
less complete ($Y>0$) and in the lower panel, clusters with
$Y\leq$0. It can be seen that there are more clusters with disk
kinematics in the lower panel.
\label{centerline}} 
\end{figure}

Figure \ref{centerline} shows these kinematics. In order to
make the expected disk signature easier to understand, we have also
plotted, for each point, the line-of-sight projected value of the disk
circular velocity if the cluster were located at the disk midplane.
For a thin disk object, this will be close to the mean of the velocity
distribution at this point, because the small disk scale height
means that no disk clusters are likely to be found very far from the
midplane. It can be seen that there are fewer globulars with disk
kinematics in the quadrant with both X and Y positive, as
expected. This is likely to be due to the lack of disk clusters there
in our current sample, as well as the possibility that the higher
reddening has made it easier to identify foreground bulge clusters
against the bright background of the bulge and disk.

\clearpage


%% If you are not including electonic art with your submission, you may
%% mark up your captions using the \figcaption command. See the 
%% No more than seven \figcaption commands are allowed per page, 
%% so if you have more than seven captions, insert a \clearpage 

%% Tables should be submitted one per page, so put a \clearpage before

%% Two options are available to the author for producing tables:  the
%% deluxetable environment provided by the AASTeX package or the LaTeX
%% table environment.  Use of deluxetable is preferred.
%%

%% Three table samples follow, two marked up in the deluxetable environment,
%% one marked up as a LaTeX table.

%% In this first example, note that the \tabletypesize{}
%% command has been used to reduce the font size of the table.
%% Note also that the \label command needs to be placed 

%% The following command ends your manuscript. LaTeX will ignore any text


\begin{thebibliography}{}

\bibitem[Adelberger et al.(2003)]{adelberger} Adelberger, K.~L.,
Steidel, C.~C., Shapley, A.~E., \& Pettini, M.\ 2003, \apj, 584, 45
\bibitem[Armandroff(1989)]{taft1989} Armandroff, T.~E.\ 1989, 
\aj, 97, 375 


\bibitem[Baade(1958)]{baade} Baade, W. 1958, in Stellar Populations,
ed. D.K.J. O'Connell (Amsterdam, North Holland), p303
\bibitem[Baade \& Arp(1964)]{baa64} Baade, W., \& Arp, H. 1964, \apj, 
139, 1027
\bibitem[Barmby et al.(2000)]{barmby00} Barmby, P., Huchra, 
J.~P., Brodie, J.~P., Forbes, D.~A., Schroder, L.~L., \& Grillmair, C.~J.\ 
2000, \aj, 119, 727 
\bibitem[Barmby, Huchra, \& Brodie(2001)]{barmbylf} Barmby, P., 
Huchra, J.~P., \& Brodie, J.~P.\ 2001, \aj, 121, 1482 
\bibitem[Barmby \& Huchra(2001)]{bh01} Barmby, P.~\& Huchra, 
J.~P.\ 2001, \aj, 122, 2458 
\bibitem[Barnes \& Hernquist(1992)]{barneshernquist} Barnes, J.~E.~\& 
Hernquist, L.\ 1992, \araa, 30, 705 
\bibitem[Battinelli, Brandimarti, \& 
Capuzzo-Dolcetta(1994)]{battinelli} Battinelli, P., Brandimarti, 
A., \& Capuzzo-Dolcetta, R.\ 1994, \aaps, 104, 379 
\bibitem[Battistini et al.(1980)]{bat80} Battistini, P., B\`onoli,
F.,Braccesi, A., Fusi Pecci, F., \& Malagnini, M. L. 1980, \aaps, 42,
357
\bibitem[Battistini et al.(1987)]{bat87} Battistini, P., B\`onoli,F.,
Braccesi, A., Federici, L., Fusi Pecci, F., Marano, B., \& B\"orngen,
F. 1987, \aaps, 67, 447
\bibitem[Battistini et al.(1993)]{bat93} Battistini, P. L.,
B\`onoli,F., Casavecchia, M., Ciotti, L., Federici, L., \& Fusi Pecci,
F. 1993, \aap, 272, 77
\bibitem[Berman \& Loinard(2002)]{bermloin} Berman, S.~\& 
Loinard, L.\ 2002, \mnras, 336, 477 
\bibitem[Binney and Tremaine(1987)]{binney} Galactic Dynamics, by
  Binney, J. and Tremaine, S., p 120, Princeton University Press,
  Princeton, New Jersey
\bibitem[Bohlin et al.(1993)]{bohlin} Bohlin, R.~C.~et al.\ 
1993, \apj, 417, 127 
\bibitem[Bottema(1993)]{bottema} Bottema, R.\ 1993, \aap, 275, 16 
\bibitem[Braun(1991)]{braun} Braun, R.\ 1991, \apj, 372, 54. 
\bibitem[Brinks \& Shane(1984)]{brinkshane} Brinks, E.~\& Shane, 
W.~W.\ 1984, \aaps, 55, 179
 \bibitem[Brodie \& Huchra(1991)]{bhuchra2} Brodie, J.~P.~\& 
Huchra, J.~P.\ 1991, \apj, 379, 157 
\bibitem[Brodie \& Huchra(1990)]{bhuchra1} Brodie, J.~P.~\& 
Huchra, J.~P.\ 1990, \apj, 362, 503 
\bibitem[Brown et al.(2003)]{tombrown} Brown, T.~M., Ferguson, 
H.~C., Smith, E., Kimble, R.~A., Sweigart, A.~V., Renzini, A., Rich, R.~M., 
\& VandenBerg, D.~A.\ 2003, astro-ph/0305318 
\bibitem[Burstein et al.(1984)]{burstein} 
Burstein, D., Faber, S.~M., Gaskell, C.~M., \& Krumm, N.\ 1984, \apj, 287, 
586 


\bibitem[Carney, Latham, \& Laird(1989)]{cll} Carney, 
B.~W., Latham, D.~W., \& Laird, J.~B.\ 1989, \aj, 97, 423
\bibitem[Chandar et al.(2002)]{chandar} 
Chandar, R., Bianchi, L., Ford, H.~C., \& Sarajedini, A.\ 2002, \apj, 564, 
712  
\bibitem[C{\^ o}t{\' e}(1999)]{cote99} C{\^ o}t{\' e}, P.\ 
1999, \aj, 118, 406 
\bibitem[Crampton et al.(1985)]{cra85} Crampton, D., Cowley, A. P.,
Schade, D., Chayer, P. 1985, \apj, 288, 494
\bibitem[Cretton et al.(2001)]{naab} 
Cretton, N., Naab, T., Rix, H., \& Burkert, A.\ 2001, \apj, 554, 291 
\bibitem[Cudworth \& Hanson(1993)]{cudhanson} Cudworth, K.~M.~\& 
Hanson, R.~B.\ 1993, \aj, 105, 168 


\bibitem[Deharveng \& Pellet(1975a)]{deharv1} Deharveng, 
J.~M.~\& Pellet, A.\ 1975, \aaps, 19, 351 
\bibitem[Deharveng \& Pellet(1975b)]{deharv2} Deharveng, 
J.~M.~\& Pellet, A.\ 1975, \aap, 38, 15 
\bibitem[Dinescu et al.(2003)]{dinescu03} Dinescu, D.~I., Girard, T.~M., van 
Altena, W.~F., \& L{\' o}pez, C.~E.\ 2003, \aj, 125, 1373 
\bibitem[Dinescu, Girard, \& van Altena(1999)]{dinescu99} 
Dinescu, D.~I., Girard, T.~M., \& van Altena, W.~F.\ 1999, \aj, 117,
1792 
\bibitem[Drimmel \& Spergel(2001)]{drimmel} Drimmel, R.~\& 
Spergel, D.~N.\ 2001, \apj, 556, 181 
\bibitem[Durrell, Harris, \& Pritchet(1994)]{durrell94} Durrell, P.~R., \
Harris, W.~E., \& Pritchet, C.~J.\ 1994, \aj, 108, 2114.
\bibitem[Durrell, Harris, \& Pritchet(2001)]{durrell01} Durrell, P.~R., \
Harris, W.~E., \& Pritchet, C.~J.\ 2001, \aj, 121, 2557.
                                            
                                       
\bibitem[Edvardsson et al.(1993)]{edvardsson} Edvardsson, B., Andersen, J\ .,
Gustafsson, B., Lambert, D.~L., Nissen, P.~E., \& Tomkin, J.\ 1993, \aap, 275,
\ 101.
\bibitem[Erb et al.(2003)]{erb} Erb, D.~K., Shapley, A.~E., 
Steidel, C.~C., Pettini, M., Adelberger, K.~L., Hunt, M.~P., Moorwood, 
A.~F.~M., \& Cuby, J.\ 2003, astro-ph/0303392 

\bibitem[Ferguson et al.(2002)]{annette} Ferguson, A.~M.~N., Irwin,
M.~J., Ibata, R.~A., Lewis, G.~F., \& Tanvir, N.~R.\ 2002, \aj, 124,
1452
\bibitem[Freeman(1987)]{kenaraa} Freeman, K.~C.\ 1987, \araa, 
25, 603 
\bibitem[Freeman(1990)]{kcf} Freeman, K.C. 1990, in ``Formation of the
  Galactic Halo .... Inside and Out'' ASP conf proc 92, Eds
  H. Morrison and A. Sarajedini
\bibitem[Frenk \& White(1982)]{frenkwhite82} Frenk, C.~S.~\& White, 
S.~D.~M.\ 1982, \mnras, 198, 173 
\bibitem[Friel(1995)]{eileenaraa} Friel, E.D. 1995, \araa, 33, 381


\bibitem[Halliday, Carter, \& Jackson(1999)]{halliday} Halliday, 
C., Carter, D., \& Jackson, Z.~C.\ 1999, Bulletin of the American 
Astronomical Society, 31, 1549 
\bibitem[Harris(1996)]{harris96} Harris, W.~E.\ 1996, \aj, 112, 
1487 
\bibitem[Huchra(1993)]{huchra} Huchra, J. 1993, in ``The Globular
Cluster-Galaxy Connection'', ASP conf series 48, eds G. Smith and J. Brodie
\bibitem[Huchra, Brodie, \& Kent(1991)]{hbrodie} Huchra, J.~P., 
Brodie, J.~P., \& Kent, S.~M.\ 1991, \apj, 370, 495 
\bibitem[Hurley-Keller et al.(2003)]{denise} Hurley-Keller, D.,
  Morrison, H.L., Harding, P. and Jacoby, G. 2003, in preparation.
\bibitem[Hurt et al.(2000)]{hurt2mass} Hurt, R.~L., Jarrett, 
T.~H., Kirkpatrick, J.~D., Cutri, R.~M., Schneider, S.~E., Skrutskie, M., 
\& van Driel, W.\ 2000, \aj, 120, 1876 

\bibitem[Ibata et al.(2001)]{ibata} Ibata, R., Irwin, M., 
Lewis, G., Ferguson, A.~M.~N., \& Tanvir, N.\ 2001, \nat, 412, 49 


\bibitem[Jenkins \& Binney(1990)]{jenkins} Jenkins, A.~\& Binney, J.\ 1990,
\mnras, 245, 305


\bibitem[Kent(1989)]{kent} Kent, S.~M.\ 1989, \aj, 97, 1614.
\bibitem[Kent, Dame, \& Fazio(1991)]{kent91} Kent, S.~M., 
Dame, T.~M., \& Fazio, G.\ 1991, \apj, 378, 131 
\bibitem[Kinman(1959)]{kinman} Kinman, T.~D.\ 1959, \mnras, 
119, 559 
\bibitem[Kissler-Patig et al.(1999)]{zepf5907} Kissler-Patig, M., Ashman, K.~M., Zepf, 
S.~E., \& Freeman, K.~C.\ 1999, \aj, 118, 197 
\bibitem[Knox, Hawkins, \& Hambly(1999)]{knox} Knox, R.~A., 
Hawkins, M.~R.~S., \& Hambly, N.~C.\ 1999, \mnras, 306, 736 
\bibitem[Kormendy(1988)]{korm88} Kormendy, J.\ 1988, \apj, 
325, 128 


\bibitem[Labbe et al.(2003)]{labbe} Labbe, I.~et al.\ 2003, 
astro-ph/0306062
\bibitem[Larson(1976)]{larson76} Larson, R.~B.\ 1976, \mnras, 
176, 31 
\bibitem[Leggett, Ruiz, \& Bergeron(1998)]{leggett} Leggett, 
S.~K., Ruiz, M.~T., \& Bergeron, P.\ 1998, \apj, 497, 294 
\bibitem[Loinard, Allen, \& Lequeux(1995)]{loin3} Loinard, 
L., Allen, R.~J., \& Lequeux, J.\ 1995, \aap, 301, 68 
\bibitem[Loinard et al.(1999)]{loinard} Loinard, L., Dame, 
T.~M., Heyer, M.~H., Lequeux, J., \& Thaddeus, P.\ 1999, \aap, 351, 1087 


\bibitem[Mateo(1998)]{marioaraa} Mateo, M.~L.\ 1998, \araa, 36, 
435 
\bibitem[Magrini et al.(2003)]{magrini} L. Magrini, L.,Corradi,
  R.L.M.,Greimel, R., Leisy, P., Lennon, D.J., Mampaso, A., Perinotto,
  M, Pollacco, D., Walsh, J., Walton, N. and Zijlstra, A. 2003, A\&A,
  in press, astro-ph/0305105 
\bibitem[Morrison(1999)]{hlm1999} Morrison, H.~L.\ 1999, ASP 
Conf.~Ser.~165: The Third Stromlo Symposium: The Galactic Halo, 174 
\bibitem[Morrison, Boroson, \& Harding(1994)]{mbh5907} 
Morrison, H.~L., Boroson, T.~A., \& Harding, P.\ 1994, \aj, 108, 1191 
\bibitem[Mould \& Kristian(1986)]{jeremy} Mould, J.~\& 
Kristian, J.\ 1986, \apj, 305, 591 
\bibitem[McElroy(1983)]{mcelroy} McElroy, D.~B.\ 1983, \apj, 
270, 485 
\bibitem[Minniti(1995)]{dante} Minniti, D.\ 1995, \aj, 109, 
1663 

\bibitem[Perrett et al.(2002)]{perrett} Perrett, K.~M., 
Bridges, T.~J., Hanes, D.~A., Irwin, M.~J., Brodie, J.~P., Carter, D., 
Huchra, J.~P., \& Watson, F.~G.\ 2002, \aj, 123, 2490
\bibitem[Peterson et al.(2003)]{ruth} Peterson, R.~C., 
Carney, B.~W., Dorman, B., Green, E.~M., Landsman, W., Liebert, J., 
O'Connell, R.~W., \& Rood, R.~T.\ 2003, \apj, 588, 299  
\bibitem[Prada Moroni \& Straniero(2002)]{moroni} Prada 
Moroni, P.~G.~\& Straniero, O.\ 2002, \apj, 581, 585 
\bibitem[Pritchet \& van den Bergh(1994)]{pvdb} Pritchet, C.~J.~\& van
den Bergh, S.\ 1994, \aj, 107, 1730.


\bibitem[Quinn \& Goodman(1986)]{quinn86} Quinn, P.~J.~\& 
Goodman, J.\ 1986, \apj, 309, 472 


\bibitem[Rao et al.(2003)]{rao}  Rao, S., Nestor, D., Turnshek, D., Lane W.,
Monier, E. and Bergeron, J. 2003, preprint, astro-ph/0211297 
\bibitem[Rich et al.(1996)]{rich96} Rich, R.~M.\,
Mighell, K.~J., Freedman, W.~L., \& Neill, J.~D.\ 1996, \aj, 111, 768.


\bibitem[Sackett et al.(1994)]{sackett5907} 
Sackett, P.~D., Morrison, H.~L., Harding, P., \& Boroson, T.~A.\ 1994, 
\nat, 370, 441 
\bibitem[Sarajedini \& Van Duyne(2001)]{ata} Sarajedini, 
A.~\& Van Duyne, J.\ 2001, \aj, 122, 2444 
\bibitem[Sargent et al.(1977)]{sar77} Sargent, W. L. W., Kowal,
C. T., Hartwick, F. D. A., \& van den Bergh, S. 1977, \aj, 82, 947
\bibitem[Schommer et al.(1991)]{schommer} Schommer, R.~A., 
Christian, C.~A., Caldwell, N., Bothun, G.~D., \& Huchra, J.\ 1991, \aj, 
101, 873 
\bibitem[Silk \& Wyse(1993)]{silkwyse} Silk, J.~\& Wyse, 
R.~F.~G.\ 1993, \physrep, 231, 295 
\bibitem[Spitzer \& Schwarzschild(1951)]{spitzer1} Spitzer, 
L.~J.~\& Schwarzschild, M.\ 1951, \apj, 114, 385 
\bibitem[Spitzer \& Schwarzschild(1953)]{spitzer2} Spitzer, 
L.~J.~\& Schwarzschild, M.\ 1953, \apj, 118, 106 
\bibitem[Shapley et al.(2003)]{shapley03} Shapley, A.~E., Steidel, C.~C., 
Pettini, M., \& Adelberger, K.~L.\ 2003, \apj, 588, 65 
\bibitem[Steidel et al.(1996)]{steidel} Steidel, C.~C., 
Giavalisco, M., Pettini, M., Dickinson, M., \& Adelberger, K.~L.\ 1996, 
\apjl, 462, L17 
\bibitem[Olszewski, Suntzeff, \& Mateo(1996)]{nickaraa} 
Olszewski, E.~W., Suntzeff, N.~B., \& Mateo, M.\ 1996, \araa, 34, 511 


\bibitem[van Dokkum \& Stanford(2001)]{vandokkum} van Dokkum, 
P.~G.~\& Stanford, S.~A.\ 2001, \apjl, 562, L35 
\bibitem[van der Kruit \& Searle(1981a)]{vdks1} van der Kruit, P.~C.~\&
Searle, L.\ 1981, \aap, 95, 105.
\bibitem[van der Kruit \& Searle(1981b)]{vdks2} van der Kruit, 
P.~C.~\& Searle, L.\ 1981, \aap, 95, 116 
\bibitem[Vete\u{s}nik(1962)]{vet62} Vete\u{s}nik, M. 1962, BAC, 13,
180     
\bibitem[Vogt et al.(1996)]{vogt} Vogt, N.~P., Forbes, 
D.~A., Phillips, A.~C., Gronwall, C., Faber, S.~M., Illingworth, G.~D., \& 
Koo, D.~C.\ 1996, \apjl, 465, L15 


\bibitem[Walker, Mihos, \& Hernquist(1996)]{walker} Walker, 
I.~R., Mihos, J.~C., \& Hernquist, L.\ 1996, \apj, 460, 121 
\bibitem[Walterbos \& Kennicutt(1988)]{wk88} Walterbos, R.~A.~M.~\&
Kennicutt, R.~C.\ 1988, \aap, 198, 61


\bibitem[Zinn(1985)]{zinn85} Zinn, R.\ 1985, \apj, 293, 424 

\end{thebibliography}
\end{document}